\documentclass[aps,ams,amsmath,prx,longbibliography,twocolumn,superscriptaddress]{revtex4-1}
\usepackage{graphics}
\usepackage{graphicx}
\usepackage{epstopdf,epsfig}
\usepackage{amsmath}
\usepackage{amssymb}
\usepackage{amsfonts}
\usepackage{mathrsfs}
\usepackage{bm}
\usepackage{xcolor}
\usepackage{bbm}
\usepackage{hyperref}
\usepackage[normalem]{ulem}
\usepackage{comment}
\usepackage{soul}
\usepackage{libertine}

\newcommand{\sgn}{\mathrm{sgn}}

\begin{document}

\title{Magneto-oscillations, nonlinearity, and nonreciprocity of Coulomb drag in quantum circuits}

\author{Alex Levchenko}
\affiliation{Department of Physics, University of Wisconsin-Madison, Madison, 53706, Wisconsin, USA}

\author{Mingyang Zheng}
\affiliation{Department of Physics, University of Florida, Gainesville, 32611, Florida, USA}
\affiliation{Department of Physics, University of Wisconsin-Madison, Madison, 53706, Wisconsin, USA}

\author{Dominique Laroche}
\affiliation{Department of Physics, University of Florida, Gainesville, 32611, Florida, USA}

\date{\today}

\begin{abstract}
We consider the problem of Coulomb drag in interactively coupled quantum circuits built of adiabatic constrictions: quantum point contacts and short quantum-wire channels. The interplay of spatial confinement and magnetic field leads to a rich oscillatory response of the drag current as a function of gate voltage and magnetic field: drag peaks track the depopulation of magnetoelectric subbands, are asymptotically periodic in inverse field, and their visibility is controlled by the competition of temperature with the field-sharpened tunneling width of the constriction. We derive a closed expression for the linear drag conductance whose interaction kernel simplifies dramatically in the experimentally relevant limit of a long thermal length compared with the range of the interwire coupling, investigate the drag in the nonlinear regime, where the drag current measures the transconductance of the drive channel at any field, and discuss physically motivated models of dissipation-induced nonreciprocity of the drag signal. Extensions accounting for Zeeman splitting, interaction renormalization of the barrier transmission, backscattering at high field, and the frequency structure of the circuit coupling delineate how each mechanism imprints itself on the temperature dependence and lineshapes of the drag oscillations.
\end{abstract}

\maketitle

%###################################################################################
%###################################################################################
%###################################################################################
%###################################################################################
%###################################################################################
\section{Introduction}\label{sec:intro}

Coulomb drag~\cite{Narozhny2016} has long served as one of the most incisive probes of electron-electron correlations in low-dimensional conductors. Because the drag signal exists only by virtue of interactions between two electrically isolated circuits, it filters out single-particle contributions that dominate ordinary transport and exposes the correlated part of the electron dynamics directly. In recent years, the reach of the technique has expanded well beyond its original setting of coupled two-dimensional electron gases~\cite{Gramila1991,Zheng1993,Jauho1993,Kamenev1995,Flensberg1995}. Drag measurements now probe quantum phases of van der Waals materials and their heterostructures, from massless carriers in graphene double layers~\cite{Kim2011,Gorbachev2012} to exciton condensation in quantum Hall bilayers and double bilayer graphene~\cite{Nandi2012,Liu2017,Li2017}. Most recently, nonreciprocal drag between Chern insulators was observed~\cite{Fu2025} as well as transport through topological surface and edge states~\cite{Du2021}. Drag provides some of the sharpest available evidence for Tomonaga-Luttinger-liquid physics in one-dimensional quantum wires~\cite{Debray2001,Yamamoto2006,Laroche2011,Laroche2014}. In many of the reported cases the observed responses do not fit comfortably within existing theoretical frameworks and call for more refined formulations. This circumstance, in part, motivates the present study.

Among the observations most in need of theoretical input are three that recur across the newest generation of one-dimensional drag experiments performed on gate-defined GaAs/AlGaAs double quantum wires, coupled both laterally and vertically~\cite{Makaju2024,Zheng2025nonlinear,Makaju2025,Zheng2026spin,Cai2026,Zheng2026}: pronounced oscillations of the drag signal as gates and magnetic field sweep the wires through their quantized-conductance structure; a strongly nonlinear dependence of the drag response on the drive bias, setting in at surprisingly small voltages; and various forms of nonreciprocity (asymmetry of the signal under reversal of the drive current or interchange of the wires) whose microscopic origin remains debated. All three phenomena are addressed in the present work within a single, deliberately minimal model: a quantum circuit of two adiabatic quantum-wire constrictions coupled solely by the Coulomb interaction.

The constriction (quantum point contact) is a uniquely convenient building block for this purpose. Its transmission is known analytically, including the orbital effect of a perpendicular magnetic field~\cite{FertigHalperin1987,Buttiker1990}, so that the entire field and gate-voltage dependence of the drag enters through a small number of exactly known energy scales. At the same time, the mechanism of drag between constrictions is qualitatively distinct from the momentum-transfer mechanism familiar from bulk double layers: it is a mesoscopic rectification effect, in which each circuit converts the nonequilibrium electric fluctuations of the other into a dc current, with an efficiency controlled by the energy dependence (the particle-hole asymmetry) of its transmission~\cite{Levchenko2008prl,Chudnovskiy2009}. This mechanism naturally produces drag peaks pinned to conductance steps, a crossover to a shot-noise-dominated nonlinear regime at bias voltages parametrically smaller than temperature, and, as we show here, a family of magneto-oscillations that mirror the magnetic depopulation of the constriction subbands.

The paper is organized as follows. Section~\ref{sec:model} recalls the saddle-point model of an adiabatic constriction and benchmarks its magnetoconductance. Section~\ref{sec:drag} presents the linear-response theory of drag between two constrictions, quotes the closed-form result valid in the experimentally relevant regime, and maps out the magneto-oscillations and their temperature and parameter dependence; the derivation, including an exact treatment of the spatial structure of the interaction kernel, is given in Appendixes~\ref{app:formalism} and \ref{app:kernel}. Section~\ref{sec:nonlinear} treats the nonlinear regime. Section~\ref{sec:nonreciprocal} develops the theory of dissipation-induced nonreciprocity of drag, building on the transmission nonreciprocity of uniformly dissipative conductors proposed in Ref.~\cite{Solow2026}. Section~\ref{sec:extensions} works out four extensions: (i) Zeeman splitting, (ii) Luttinger-liquid renormalization of the constriction transmission, (iii) the high-field backscattering channel, and (iv) the frequency structure of the circuit coupling. Section~\ref{sec:summary} summarizes the results and discusses their generality and their connection to ongoing experiments.

%###################################################################################
%###################################################################################
%###################################################################################
%###################################################################################
%###################################################################################
\section{Adiabatic constriction in a magnetic field}\label{sec:model}

\begin{figure*}[t]
\begin{minipage}{0.24\textwidth}\textbf{(a)}\\ \includegraphics[width=\linewidth]{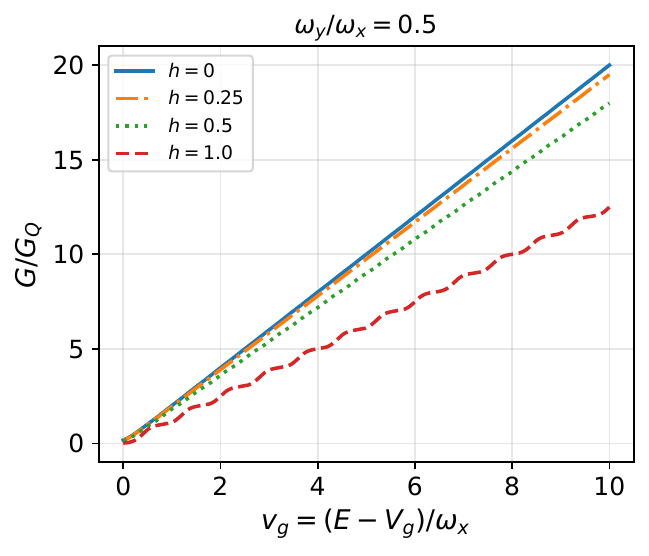}\end{minipage}\hfill
\begin{minipage}{0.24\textwidth}\textbf{(b)}\\ \includegraphics[width=\linewidth]{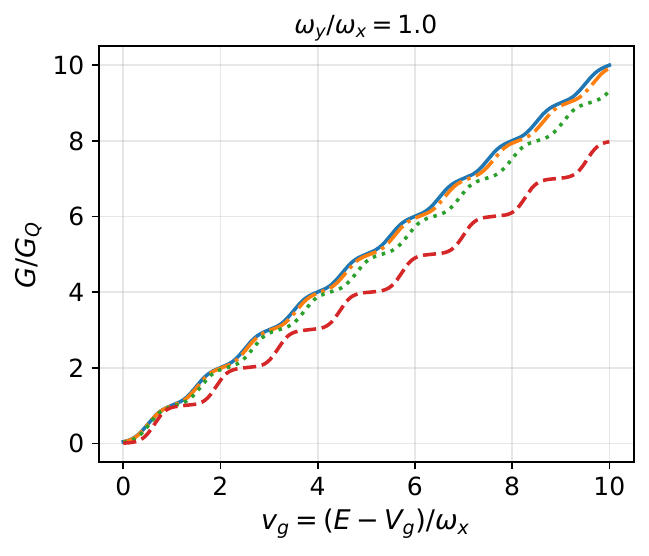}\end{minipage}\hfill
\begin{minipage}{0.24\textwidth}\textbf{(c)}\\ \includegraphics[width=\linewidth]{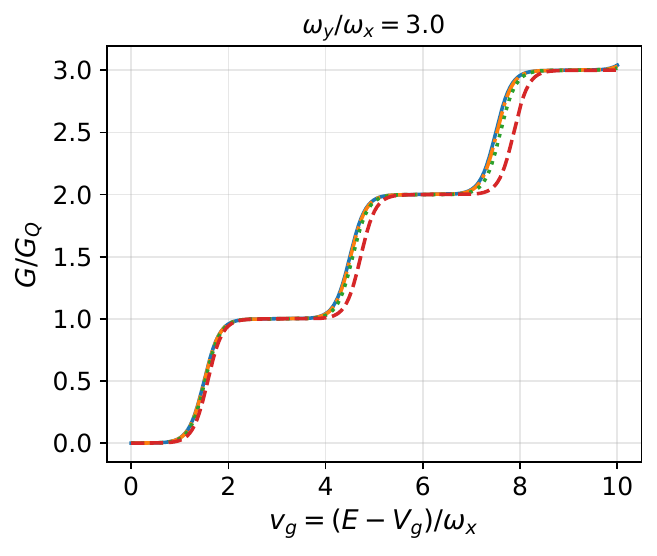}\end{minipage}\hfill
\begin{minipage}{0.24\textwidth}\textbf{(d)}\\ \includegraphics[width=\linewidth]{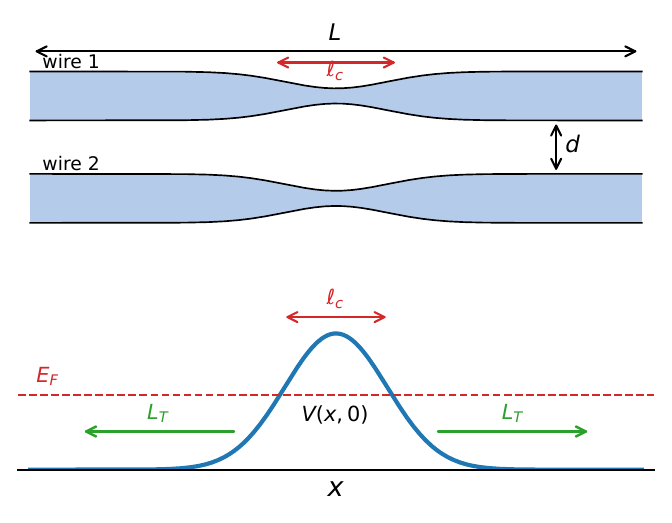}\end{minipage}
\caption{Magnetoconductance of the saddle-point constriction, Eqs.~\eqref{eq:kemble}--\eqref{eq:landauer}, for confinement ratios (a) $\omega_y/\omega_x=1/2$, (b) $1$, and (c) $3$, at several values of the reduced field $h=\omega_c/\omega_x$. The staircase sharpens and the plateaus widen with increasing field as $\Delta_1(B)$ shrinks and $\Delta_2(B)$ grows, reproducing the classic results of Refs.~\cite{FertigHalperin1987,Buttiker1990} and benchmarking the model used throughout. (d) Schematic of the coupled double-wire geometry: gates define two parallel channels of length $L$ at separation $d$; the potential profile along each channel has its bottleneck (extent $\ell_c\ll L$) described by the saddle Eq.~\eqref{eq:saddle}, while the interwire coupling acts along the entire parallel region, over which the rectified perturbations spread ballistically within $L_T$.}
\label{fig:conductance}
\end{figure*}

We model each constriction by the adiabatic saddle-point potential [see Fig.~\ref{fig:conductance}(d) for the sketch]
\begin{equation}
V(x,y)=V_g-\tfrac12 m\omega_x^2x^2+\tfrac12 m\omega_y^2y^2,
\label{eq:saddle}
\end{equation}
where the $x$ axis runs along the transport direction, $V_g$ is the electrostatic potential at the saddle, and the curvatures of the potential define the longitudinal and transverse frequencies $\omega_x$ and $\omega_y$. All three parameters are the coefficients of the quadratic expansion of the electrostatic potential about the bottleneck and are set jointly by the gate voltages, the device geometry, and the density of the two-dimensional gas; a gate sweep moves primarily $V_g$ while $\omega_{x,y}$ drift slowly, so the theory below treats them as independent coordinates of a parameter plane that a physical sweep traverses along a diagonal-like path. At zero magnetic field the problem separates into transverse channels $n=0,1,2,\dots$, and transmission through the resulting inverted parabolic barrier is given by the Kemble-Connor formula~\cite{Kemble1935,Connor1968,Glazman1988}
\begin{equation}
T_n(E)=\frac{1}{1+e^{-\pi\varepsilon_n}},
\qquad
\varepsilon_n=\frac{E-V_g-\Delta_2\,(n+\tfrac12)}{\Delta_1},
\label{eq:kemble}
\end{equation}
with $\Delta_1=\omega_x/2$ and $\Delta_2=\omega_y$ (natural units $\hbar=k_B=1$ are used throughout). A perpendicular magnetic field $B$ preserves the quadratic form of the Hamiltonian, and the exact single-particle solution of Fertig and Halperin~\cite{FertigHalperin1987}, cast by B\"uttiker~\cite{Buttiker1990} in the form of Eq.~\eqref{eq:kemble}, amounts to the replacement of the two energy scales by
\begin{align}
\Delta_1(B)&=\frac{1}{2\sqrt2}\Big[\sqrt{\Omega^4+4\omega_x^2\omega_y^2}-\Omega^2\Big]^{1/2},
\nonumber\\
\Delta_2(B)&=\frac{1}{\sqrt2}\Big[\sqrt{\Omega^4+4\omega_x^2\omega_y^2}+\Omega^2\Big]^{1/2},
\label{eq:delta12}
\end{align}
where $\Omega^2=\omega_c^2+\omega_y^2-\omega_x^2$ and $\omega_c=eB/m$ is the cyclotron frequency. The physical content of Eq.~\eqref{eq:delta12} is transparent in the limits. At $B=0$ one recovers $\Delta_1=\omega_x/2$, $\Delta_2=\omega_y$. At strong fields, $\omega_c\gg\omega_{x,y}$, the level spacing approaches the magnetoelectric-subband value $\Delta_2\to\sqrt{\omega_c^2+\omega_y^2-\omega_x^2}\simeq\omega_c$, while the tunneling width collapses, $\Delta_1\to\omega_x\omega_y/2\omega_c$: the magnetic field simultaneously stretches the staircase and sharpens its conductance steps. Throughout we use the dimensionless field and gate variables
\begin{equation}
h=\frac{\omega_c}{\omega_x},\qquad
a=\frac{\omega_y}{\omega_x},\qquad
v_{g,i}=\frac{E_F-V_{g,i}}{\omega_x},
\end{equation}
for wire $i=1,2$. For GaAs parameters ($m^*=0.067m_e$) the conversion is $h\simeq 1.73\,B[\mathrm{T}]/\omega_x[\mathrm{meV}]$, so that for a typical $\omega_x\sim1$--2\,meV the interesting range $h\sim1$--3 corresponds to laboratory fields of a few tesla.

The linear conductance follows from the Landauer formula,
\begin{equation}
G=G_Q\sum_n T_n(E_F),\qquad G_Q=\frac{e^2}{\pi},
\label{eq:landauer}
\end{equation}
and is shown in Fig.~\ref{fig:conductance} for three confinement ratios. The evolution of the staircase with field (plateau widening, plateau transitions, and the depopulation of channels) reproduces the classic magnetoconductance results~\cite{Buttiker1990,Berggren1986,vanWees1988B} and serves as the benchmark of the model: every energy scale that will control the drag below is already visible in these curves.

Two remarks connect the model to the gate-defined devices. First, on geometry: the gates define the full one-dimensional channel of length $L$ (of order a micron. Along this channel the electrostatic potential is not flat; its softest point, the bottleneck near pinch-off, is what the saddle Eq.~\eqref{eq:saddle} describes, with $\ell_c\sim(m\omega_x)^{-1/2}\ll L$ the extent of the curvature region [see Fig.~\ref{fig:conductance}(d)]. The transport coefficients $T_n$ are controlled by the bottleneck, while the interwire Coulomb coupling of Sec.~\ref{sec:drag} acts along the entire parallel double-wire region, because the particle-hole-asymmetric disturbances created at the bottleneck propagate ballistically over the thermal length $L_T=v_F/T$ into the channel on both sides. The local gating thus creates both the wire and, through its potential profile, the effective constriction. The two constrictions need not be identical, and their bottlenecks need not face each other: a longitudinal offset $x_0$ between them leaves the drag unaffected as long as $x_0\ll L_T$, while supplying a coupling asymmetry that is quantified in Appendix~\ref{app:kernel}. Second, on symmetry. A left/right-asymmetric saddle, e.g., one with a cubic correction $\lambda x^3$, arises whenever the gates of the two wires are not swept together, which is generic in vertically coupled devices. Such an asymmetry leaves the Kemble form of transmission intact at leading order. It is, however, precisely what generates the circuit inversion asymmetry $\alpha_-$ of Sec.~\ref{sec:nonlinear}, the transmission-time difference $\Delta\tau$ of Sec.~\ref{sec:nonreciprocal}, and the coupling asymmetry $\eta_c$. The nonlinear and nonreciprocal effects discussed below should therefore be regarded as expected rather than exceptional in such devices.

%###################################################################################
%###################################################################################
%###################################################################################
%###################################################################################
%###################################################################################
\section{Linear Coulomb drag between two constrictions}\label{sec:drag}

\subsection{Rectification mechanism and the drag formula}\label{sec:dragformula}

We now couple two such constrictions by the Coulomb interaction while keeping them electrically isolated, in the geometry of the quantum-circuit drag problem~\cite{Levchenko2008prl} realized experimentally in coupled point contacts~\cite{Khrapai2007} and in laterally and vertically coupled quantum wires~\cite{Laroche2011,Laroche2014,Makaju2024}. A weak drive current in circuit~1 generates nonequilibrium potential fluctuations that act on circuit~2; because the transmissions $T_n(\varepsilon)$ of circuit~2 depend on energy, the electron and hole excitations created by these fluctuations are transmitted with different probabilities, and a net dc drag current results. The efficiency of this rectification is quantified by the particle-hole asymmetry factor of each wire,
\begin{equation}
A_i(\omega)=\int d\varepsilon\,
\big[f(\varepsilon_-)-f(\varepsilon_+)\big]
\sum_n\big[T^{(i)}_n(\varepsilon_+)-T^{(i)}_n(\varepsilon_-)\big],
\label{eq:Afactor}
\end{equation}
where $\varepsilon_\pm=\varepsilon\pm\omega/2$, $f$ is the Fermi function, and $\omega$ is the frequency of the interwire potential fluctuation being rectified. The asymmetry is largest when the Fermi level of a wire sits at a conductance step, where the energy dependence of $T_n$ is strongest; on the plateaus, $A_i$ is exponentially small. It is this simple observation that underlies all of the oscillation physics below~\cite{Levchenko2008prl,Khrapai2007}.

The linear drag conductance is obtained from a Keldysh calculation with the scattering states of the two constrictions (Appendix~\ref{app:formalism}). It takes the form
\begin{equation}
g_D=\frac{1}{8\pi T}\left(\frac{e}{8\pi v_F^2}\right)^{\!2}\!\int\limits_{-\infty}^{+\infty}\!\!
\frac{\mathcal J\!(k) d\omega}{\sinh^2(\omega/2T)}
A_1(\omega)A_2(\omega),
\label{eq:master}
\end{equation}
where the kernel $\mathcal J(k)$ with $k=\omega/v_F$ carries the entire spatial structure of the interwire interaction $\mathcal V(x_1,x_2)$. A central technical result of this work, derived in Appendix~\ref{app:kernel}, is that the four spatial integrals defining $\mathcal J$ factorize exactly (with no assumption about the range or smoothness of $\mathcal V$) into half-line Fourier transforms of the interaction,
\begin{align}
\mathcal J(k)&=\sum_{s,s'=\pm} s\,s'\,\big|V_{ss'}(k)\big|^2,
\nonumber\\
V_{ss'}(k)&=\int\limits_{sx>0}\!dx\!\int\limits_{s'y>0}\!dy\;
e^{ik(sx+s'y)}\,\mathcal V(x,y),
\label{eq:kernel}
\end{align}
with $s$ ($s'$) labeling the side of constriction 1 (2). The drag thus measures the coherent same-side minus cross-side coupling of the particle-hole-asymmetric density disturbances emitted by the two constrictions at wavevector $k=\omega/v_F\sim L_T^{-1}$, with $L_T=v_F/T$ the ballistic thermal length. The derivation retains only the smooth (non-$2k_F$) components of the densities; the momentum-transfer (backscattering) channel thereby discarded is restored and estimated in Sec.~\ref{sec:edge}. Equation~\eqref{eq:kernel} passes two nontrivial checks: a spatially uniform $\mathcal V$ gives $\mathcal J\equiv0$, as gauge invariance demands, and a lead-resolved (capacitive) coupling reproduces exactly the trans-impedance combination $|Z_{LL}|^2+|Z_{RR}|^2-2Z_{LR}Z_{RL}$ of the circuit theory of Ref.~\cite{Levchenko2008prl}. We also note that $\mathcal J$ is not sign-definite, a point we return to in the summary in connection with the negative drag observed experimentally~\cite{Yamamoto2006,Laroche2011}.

A striking simplification occurs in the limit most relevant to experiments. For wires coupled over a long window by a translationally invariant interaction of range $d$ (set by the interwire distance and gate screening), and at temperatures such that $L_T\gg d$, the kernel collapses onto the zero-momentum component of the interaction, $\langle\mathcal J\rangle=\hat{\mathcal V}(0)^2/k^2$ with $\hat{\mathcal V}(0)=\int\mathcal V(u)\,du$ (Appendix~\ref{app:kernel}). For $v_F/d\simeq1$--4\,meV ($\sim$10--50\,K) this condition corresponds to the operating regime of most experiments.

This collapse is what distinguishes the rectification mechanism from the momentum-transfer drag of bulk double layers~\cite{Gramila1991,Zheng1993,Jauho1993,Kamenev1995,Flensberg1995}, where the kernel is the dynamically screened interaction at wavevectors $q\sim k_F$ and plasmon enhancement and screening are decisive. Here the wavevector transferred is the thermal one, $k\sim L_T^{-1}\ll k_F$, and the interwire coupling enters only through the single number $\hat{\mathcal V}(0)$: its shape, momentum dependence, and screening affect the overall amplitude of the drag but none of the lineshapes, field, gate, or temperature dependences derived below. Screening enters $\hat{\mathcal V}(0)$ at two levels. The electrostatic screening by the gates is part of $\mathcal V$ itself and is what renders $\hat{\mathcal V}(0)$ finite: for two wires at separation $d$ beneath a metallic gate at height $h_g$, the image-charge form $\mathcal V(u)=(e^2/\epsilon)[(u^2+d^2)^{-1/2}-(u^2+d^2+4h_g^2)^{-1/2}]$ gives $\hat{\mathcal V}(0)=(2e^2/\epsilon)\ln(\sqrt{d^2+4h_g^2}/d)\simeq(2e^2/\epsilon)\ln(2h_g/d)$, i.e., the logarithmic dependence on the wire separation with screening length $\lambda_s=2h_g$ invoked in Sec.~\ref{sec:edge} and Appendix~\ref{app:kernel}, with $\hat{\mathcal F}(0)=\ln(1+4h_g^2/d^2)\approx2$--$3.5$ for $d=50$--$100$\,nm and $h_g=100$--$150$\,nm. The screening by the carriers of the wires themselves (the intra- and interwire polarization that defines the Luttinger parameters of the coupled pair) renormalizes $\hat{\mathcal V}(0)$ multiplicatively; since the theory is of second order in the interwire coupling, this renormalization is absorbed in the one overall amplitude. Only when $L_T\lesssim d$, i.e., above $v_F/d$, does the shape of $\mathcal V$ reenter, through the value $\mathcal V(0)\sim e^2/\epsilon d$ at zero separation along the wire (Appendix~\ref{app:kernel}).

In the limit $L_T\gg d$ the drag conductance acquires the closed form
\begin{align}
\frac{g_D}{g_Q}&=\frac{\pi^2\hat{\mathcal F}^2(0)}{3072}\,r_s^2\,
\frac{T^2}{\Delta_1^{(1)}\Delta_1^{(2)}}\, \nonumber \\
&\times\prod_{i=1,2}\;\sum_n
\cosh^{-2}\!\left[\frac{\pi}{2}\,\frac{v_{g,i}-\Delta_2^{(i)}(B)(n+\tfrac12)/\omega_x}{\Delta_1^{(i)}(B)/\omega_x}\right],
\label{eq:gdmain}
\end{align}
valid for $\pi T\ll\Delta_1^{(i)}$, where $r_s=e^2/v_F$ is the gas parameter, $\mathcal V=e^2\mathcal F$ defines the dimensionless interaction profile, and $\Delta_1^{(i)}(B)$ and $\Delta_2^{(i)}(B)$ are the tunneling width and the subband spacing of constriction $i$, each given by Eq.~\eqref{eq:delta12} with the saddle parameters $(\omega_x^{(i)},\omega_y^{(i)})$ of that constriction; the effective width of its transmission step is $\Delta_1^{(i)}(B)/\pi$ [viewed as a function of energy or, equivalently, gate voltage, the saddle transmission Eq.~\eqref{eq:kemble} has the mathematical form of a Fermi-Dirac step of this width]. The two constrictions may thus have different confinements; equal confinement, $\Delta_{1,2}^{(1)}=\Delta_{1,2}^{(2)}$, is assumed in the figures only for simplicity. Unequal confinement displaces the two resonance combs of Eq.~\eqref{eq:gdmain} relative to each other and lowers the oscillation contrast without changing the lineshape of either comb. In the opposite, thermally smeared regime $\pi T\gg\Delta_1^{(i)}$ the same expression holds with $\Delta_1^{(i)}/\pi\to T$, which removes the prefactor $\pi^2T^2/\Delta_1^{(1)}\Delta_1^{(2)}$ and yields
\begin{equation}
\frac{g_D}{g_Q}=\frac{\hat{\mathcal F}^2(0)}{3072}\,r_s^2
\prod_{i=1,2}\sum_n
\cosh^{-2}\!\left[\frac{v_{g,i}\,\omega_x-\Delta_2^{(i)}(n+\tfrac12)}{2T}\right]:
\label{eq:gdhighT}
\end{equation}
the drag at the peaks is then temperature independent, while the comb width grows as $T$. Equation~\eqref{eq:gdmain} contains the complete gate, field, and temperature dependence of the linear drag in terms of exactly known functions: the drag is a product of two ``resonance combs,'' one per wire, with teeth of width $\max(\Delta_1^{(i)}(B)/\pi,\,T)$ centered on the steps $v_{g,i}=\Delta_2^{(i)}(B)(n+\tfrac12)/\omega_x$, riding on the prefactor $\pi^2T^2/\Delta_1^{(1)}(B)\Delta_1^{(2)}(B)$ that grows as the field compresses the steps. Equation~\eqref{eq:gdmain} also fixes the absolute scale of the effect. With $r_s=e^2/\epsilon v_F\approx1.7$ for GaAs ($\epsilon=12.9$, $v_F=1.5\times10^5$\,m/s) and $\hat{\mathcal F}(0)\approx2.5$, the prefactor $\pi^2\hat{\mathcal F}^2(0)r_s^2/3072\approx0.06$; at a peak, for $T/\Delta_1=0.1$ (e.g., $\Delta_1=0.5$\,meV at $T\approx0.6$\,K), this gives $g_D\approx6\times10^{-4}g_Q\approx45$\,nS, i.e., a drag resistance $R_D=V_{\rm drag}/I_{\rm drive}\simeq g_D/G_1G_2$ of order $10\,\Omega$ for single-channel wires, saturating in the thermally smeared regime of Eq.~\eqref{eq:gdhighT} at $g_D/g_Q\approx6\times10^{-3}$, $R_D\sim10^2\,\Omega$. These are the orders of magnitude reported for gate-defined double wires~\cite{Laroche2011,Laroche2014,Makaju2024}.

\subsection{Magneto-oscillations}\label{sec:oscillations}

\begin{figure*}[t]
\begin{minipage}{0.32\textwidth}\textbf{(a)}\\ \includegraphics[width=\linewidth]{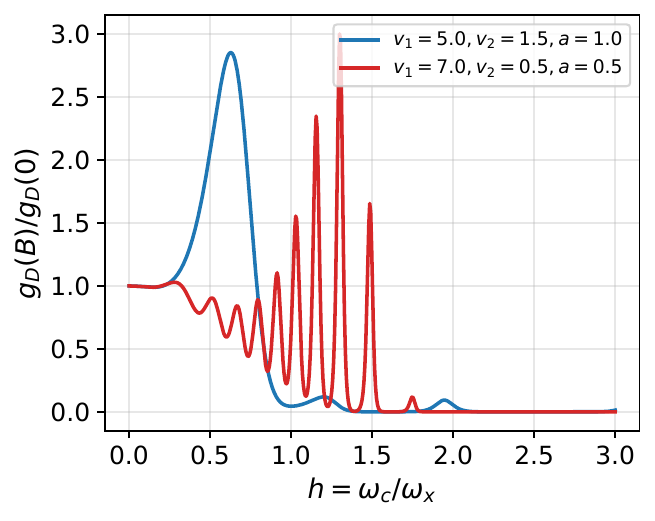}\end{minipage}\hfill
\begin{minipage}{0.32\textwidth}\textbf{(b)}\\ \includegraphics[width=\linewidth]{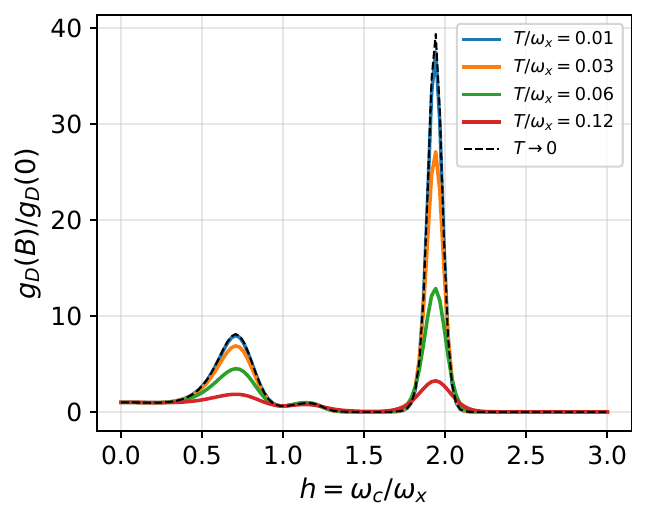}\end{minipage}\hfill
\begin{minipage}{0.32\textwidth}\textbf{(c)}\\ \includegraphics[width=\linewidth]{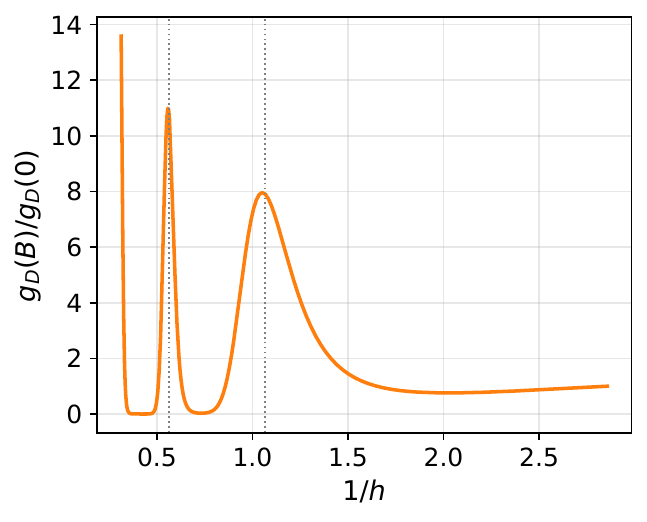}\end{minipage}
\caption{Magneto-oscillations of the linear drag. (a) Normalized drag $g_D(B)/g_D(0)$ from Eq.~\eqref{eq:gdmain} for two representative parameter sets $(v_{g,1},v_{g,2},a)$. (b) Thermal smearing at $v_{g,1}=5$, $v_{g,2}=3$, $a=1$: exact finite-$T$ evaluation of Eqs.~\eqref{eq:Afactor} and \eqref{eq:master} for increasing $T/\omega_x$; the $T\to0$ limit Eq.~\eqref{eq:gdmain} is shown dashed. (c) Matched wires ($v_{g,1}=v_{g,2}=5$, $a=1.2$, $T/\omega_x=0.1$) plotted against $1/h$: the dotted lines mark the depopulation condition $\Delta_2(h)(n+\tfrac12)=v_{g,1}\omega_x$, demonstrating the asymptotic $1/B$ periodicity.}
\label{fig:oscillations}
\end{figure*}

\begin{figure*}[t]
\begin{minipage}{0.32\textwidth}\textbf{(a)}\\ \includegraphics[width=\linewidth]{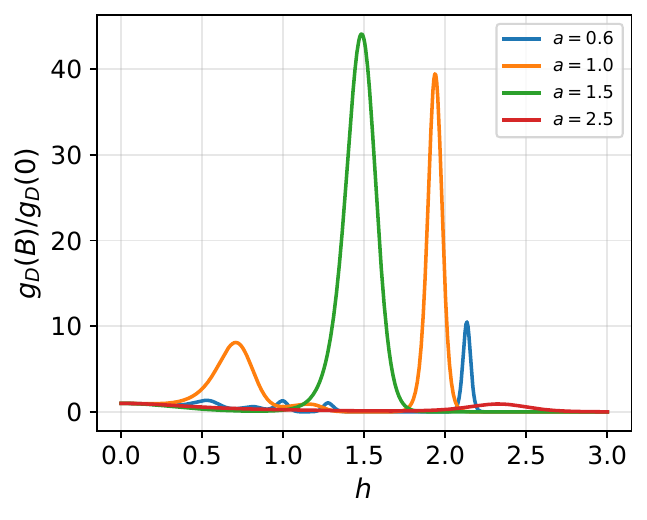}\end{minipage}\hfill
\begin{minipage}{0.32\textwidth}\textbf{(b)}\\ \includegraphics[width=\linewidth]{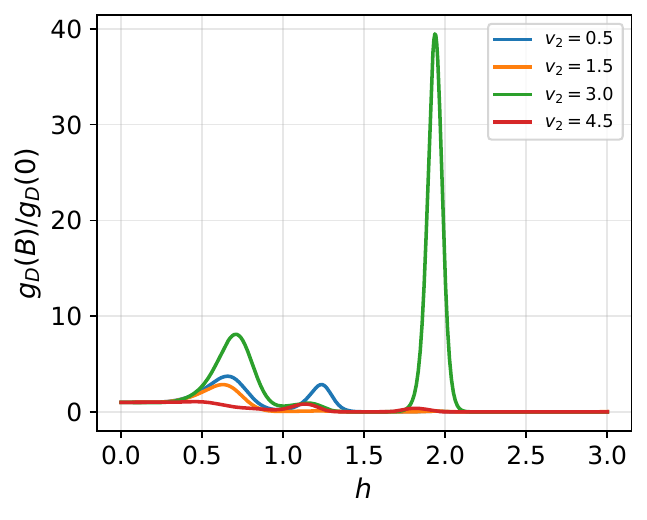}\end{minipage}\hfill
\begin{minipage}{0.32\textwidth}\textbf{(c)}\\ \includegraphics[width=\linewidth]{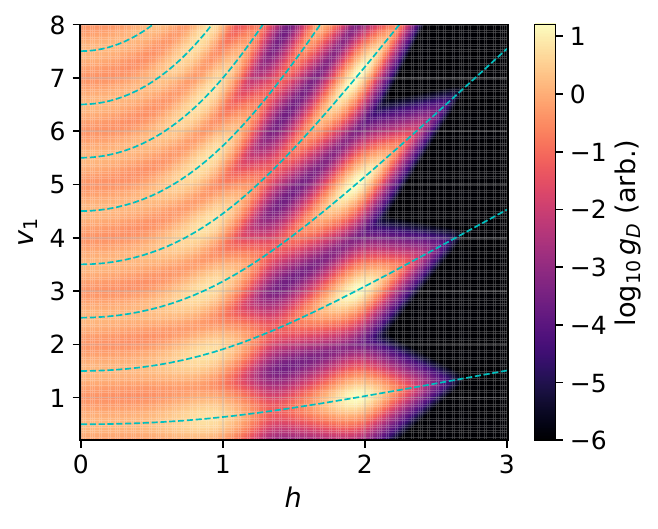}\end{minipage}
\caption{Parameter dependence of the drag magneto-oscillations, Eq.~\eqref{eq:gdmain}. (a) Confinement-ratio dependence at fixed gates ($v_{g,1}=5$, $v_{g,2}=3$). (b) Drag-wire gate dependence at $v_{g,1}=5$, $a=1$. (c) Map of $\log_{10}g_D$ in the $(h,v_{g,1})$ plane at $v_{g,2}=3$, $a=1$; the dashed curves are the subband risers $\Delta_2(h)(n+\tfrac12)$, which the drag ridges track. This map is the direct theoretical counterpart of the experimental drag-versus-gate-and-field color plots. In a physical gate sweep $v_{g,i}$ and $a$ vary together (Sec.~\ref{sec:model}); the panels scan them independently to disentangle their effects.}
\label{fig:params}
\end{figure*}

Figure~\ref{fig:oscillations} presents the resulting magneto-oscillations. Their shape follows directly from Eq.~\eqref{eq:gdmain}: the drag peaks whenever a magnetoelectric subband step of either wire crosses its Fermi level, $v_{g,i}=\Delta_2(h)(n+\tfrac12)/\omega_x$. These are precisely the fields at which the conductance of that wire steps down by one quantum (the magnetic-depopulation fields) and at large $h$, where $\Delta_2\simeq\omega_c$, the peak positions become periodic in $1/B$ [Fig.~\ref{fig:oscillations}(c)], in direct analogy with the Shubnikov--de Haas effect. Between the peaks the drag is exponentially small, so the zero-temperature theory over-resolves the oscillations: the experimentally observed contrast is set by the competition between $T$ and the field-sharpened step width $\Delta_1(B)/\pi$. Panel \ref{fig:oscillations}(b) quantifies this with the exact finite-temperature evaluation of Eqs.~\eqref{eq:Afactor} and \eqref{eq:master}, which interpolates automatically between the two limits of Eq.~\eqref{eq:gdmain}: as $T/\omega_x$ grows from $0.01$ to $0.12$ the oscillation contrast collapses from two orders of magnitude to a factor of a few, which is the regime seen experimentally. For GaAs parameters the realistic window is $T/\omega_x\sim0.01$--$0.07$. Note that the normalization by $g_D(0)$ hides the overall growth of the absolute drag with temperature, $g_D\propto T^2$ in this regime: raising $T$ reduces the oscillation contrast while increasing the signal itself [cf.\ Figs.~\ref{fig:myg}(a), \ref{fig:circuit}(a), and \ref{fig:kernel}(b)].

Figure~\ref{fig:params} maps the parameter space. Softer transverse confinement (smaller $a$) packs more channels below the Fermi level and produces more oscillations before the wire empties; the drag-wire gate voltage $v_{g,2}$ selects which of its steps participate; and matched wires ($v_{g,1}=v_{g,2}$), the natural configuration of nominally symmetric double-wire devices, give the largest contrast because the resonance combs of the two wires coincide. The $(h,v_{g,1})$ map of Fig.~\ref{fig:params}(c) is the most compact summary: the drag ridges track the subband steps of the swept wire, interrupted where the fixed wire is off resonance. Overlaying such a map with the measured transconductance of each wire provides a parameter-free test of the rectification mechanism: in the sense of Eqs.~\eqref{eq:gdmain} and \eqref{eq:transcond}, no adjustable parameters enter beyond one overall amplitude.

%###################################################################################
%###################################################################################
%###################################################################################
%###################################################################################
%###################################################################################
\section{Nonlinear regime}\label{sec:nonlinear}

\begin{figure*}[t]
\begin{minipage}{0.32\textwidth}\textbf{(a)}\\ \includegraphics[width=\linewidth]{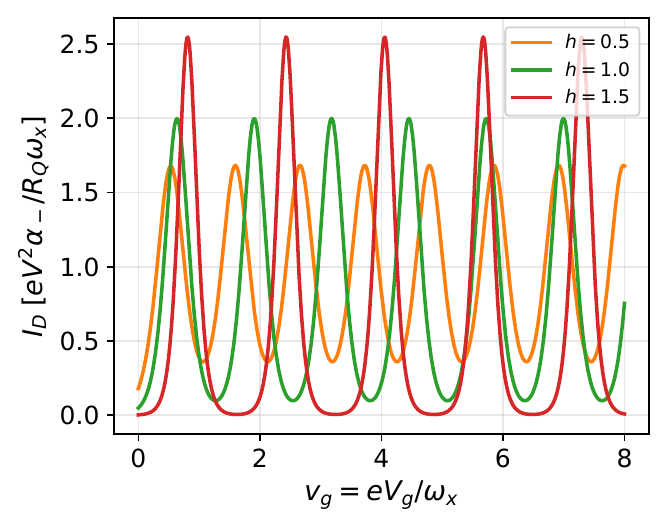}\end{minipage}\hfill
\begin{minipage}{0.32\textwidth}\textbf{(b)}\\ \includegraphics[width=\linewidth]{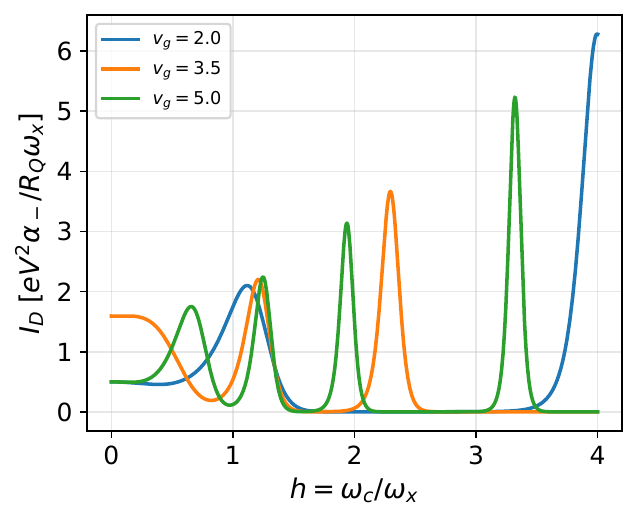}\end{minipage}\hfill
\begin{minipage}{0.32\textwidth}\textbf{(c)}\\ \includegraphics[width=\linewidth]{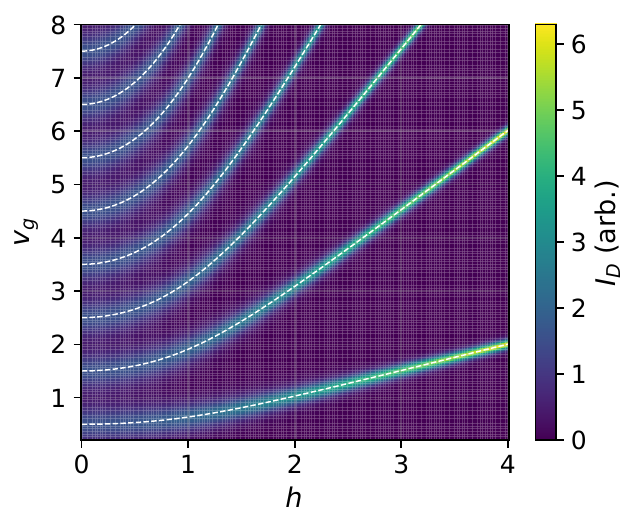}\end{minipage}
\caption{Nonlinear drag in a magnetic field, Eq.~\eqref{eq:transcond}, in units of $eV^2\alpha_-/R_Q\omega_x$ ($a=1$). (a) Gate dependence for several fields: the peaks sharpen and grow $\propto1/\Delta_1(h)$ as the field compresses the risers. (b) Magneto-oscillations at fixed gate: each peak marks one subband depopulation, and the last (highest-field) peak is the strongest. (c) Map of $I_D(v_g,h)$; the signal concentrates on the riser fan $\Delta_2(h)(n+\tfrac12)$ (dashed).}
\label{fig:nonlinear}
\end{figure*}

At larger drive bias the drag ceases to be linear in $V$. The crossover was identified in Ref.~\cite{Levchenko2008prl}: for $T\ll\Delta_1$ it occurs at the parametrically small voltage $eV^*\simeq T^2/\Delta_2^{\rm c}=\pi T^2/\Delta_1\ll T$, beyond which the drag current is no longer a rectification of near-equilibrium thermal fluctuations but of the quantum shot noise of the drive circuit~\cite{Lesovik1989,Reznikov1995,BlanterButtiker2000,Khrapai2007,Aguado2000,Onac2006}. In this regime
\begin{equation}
I_D=\frac{eV^2}{\Delta_2^{\rm c}R_Q}\,\alpha_-(0)\sum_n T_n(1-T_n),
\label{eq:nlin}
\end{equation}
where $R_Q=2\pi/e^2$, $\alpha_-(0)$ is the inversion-asymmetry parameter of the electrostatic environment, and $\Delta_2^{\rm c}$ is the curvature scale of the drive constriction, related to the saddle parameters by $\Delta_2^{\rm c}=\Delta_1/\pi$. The generalization to a finite magnetic field can be argued as follows: nothing in the derivation of Eq.~\eqref{eq:nlin} (the two-terminal scattering-theory noise formula and the electrostatic kernel $\alpha_-$) invokes time-reversal symmetry, so the field enters only through the transmissions $T_n\to T_n(B)$ of Eqs.~\eqref{eq:kemble}--\eqref{eq:delta12} and through $\Delta_2^{\rm c}\to\Delta_1(B)/\pi$. Because the saddle transmission has the form of a Fermi--Dirac step in gate voltage, of width $\Delta_1(B)/\pi$, one has the exact identity $T_n(1-T_n)=(\Delta_1/\pi)\,\partial T_n/\partial(eV_g)$, and Eq.~\eqref{eq:nlin} collapses to
\begin{equation}
I_D(V_g,B)=\frac{eV^2}{R_Q}\,\alpha_-(0)\,
\frac{\partial}{\partial (eV_g)}\!\left(\frac{G_1(V_g,B)}{G_Q}\right).
\label{eq:transcond}
\end{equation}
Within this theory the nonlinear drag magneto-oscillations are identical in shape to the transconductance of the drive channel alone, at any field. This is the sharpest experimentally testable statement of the shot-noise mechanism: the measured nonlinear drag map and the separately measured $dG_1/dV_g$ map must coincide up to one global constant. Note also that Eq.~\eqref{eq:transcond} carries no explicit temperature dependence: within its validity window the nonlinear drag is $T$ independent in this mechanism.

Figure~\ref{fig:nonlinear} displays the consequences. The peaks of $\sum_nT_n(1-T_n)$, the Fano factor of the drive constriction~\cite{BlanterButtiker2000}, sit at the steps, and their amplitude grows with field as $1/\Delta_1(B)\propto B$, in sharp contrast to the linear regime where thermal smearing caps the peak height once $\Delta_1(B)\lesssim\pi T$. Opposite envelope behavior of the oscillations, growing with field in the nonlinear regime, saturating or decaying in the linear one, is thus a fingerprint that distinguishes shot-noise drag from thermal-rectification drag. The validity window of Eq.~\eqref{eq:transcond}, $\pi T^2/\Delta_1(B)$$\ll eV\lesssim\Delta_1(B)$, closes at the field where $\Delta_1(B)\sim T$. Since the crossover voltage $eV^*(B)=\pi T^2/\Delta_1(B)$ grows with field, a bias that is safely nonlinear at $B=0$ can drift back into the linear regime at high field. The drag pattern then converts from the single-wire transconductance fan of Eq.~\eqref{eq:transcond} to the two-wire product form of Eq.~\eqref{eq:gdmain}. Observing this conversion would strongly corroborate the theory. We note that the nonlinear crossover at anomalously small bias, the $V^2$ scaling, and the sensitivity of the drag sign to the drive polarity have all been reported in the double-wire devices of Refs.~\cite{Zheng2025nonlinear,Makaju2025,Zheng2026spin}.

%###################################################################################
%###################################################################################
%###################################################################################
%###################################################################################
%###################################################################################
\section{Nonreciprocity from dissipation}\label{sec:nonreciprocal}

\begin{figure*}[t]
\begin{minipage}{0.32\textwidth}\textbf{(a)}\\ \includegraphics[width=\linewidth]{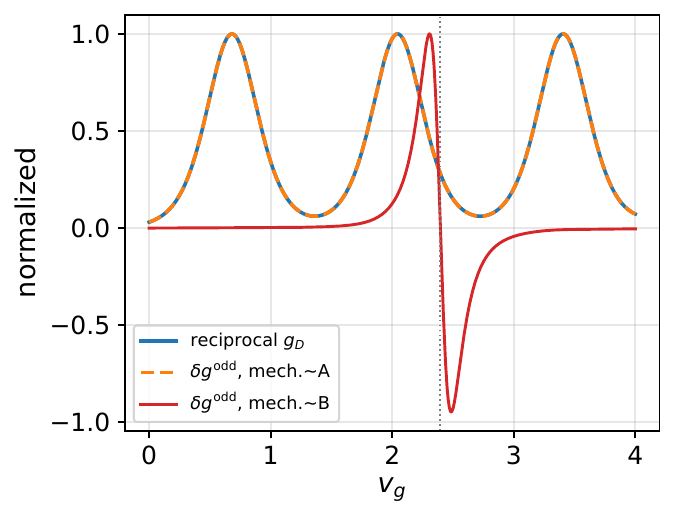}\end{minipage}\hfill
\begin{minipage}{0.32\textwidth}\textbf{(b)}\\ \includegraphics[width=\linewidth]{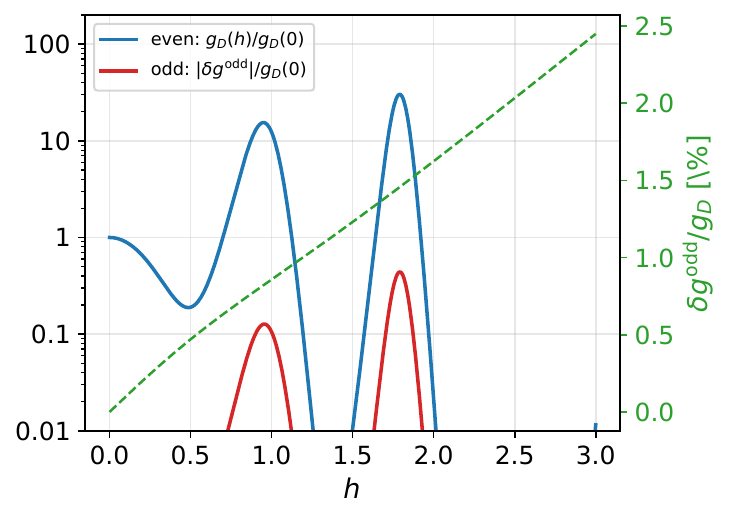}\end{minipage}\hfill
\begin{minipage}{0.32\textwidth}\textbf{(c)}\\ \includegraphics[width=\linewidth]{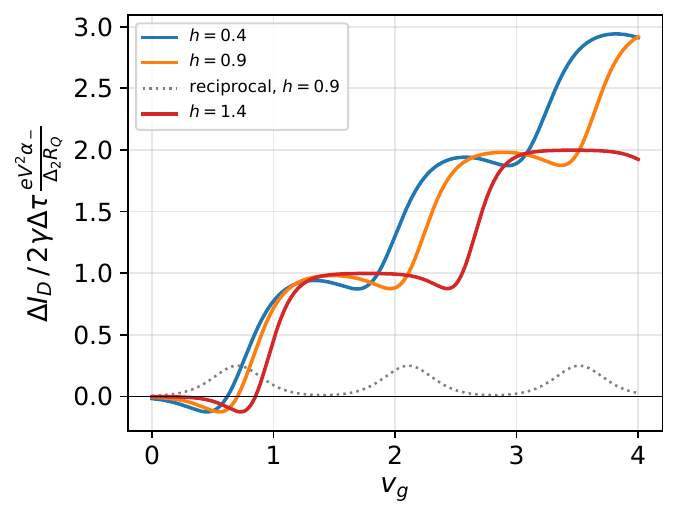}\end{minipage}
\caption{Dissipation-induced nonreciprocity of drag. (a) Lineshapes of the field-odd component of the linear drag at $h=0.8$, $a=1.2$: for an energy-smooth transmission-time difference (mechanism A) $\delta g^{\rm odd}$ tracks the reciprocal peaks, while a resonant $\Delta\tau$ at a subband anticrossing (mechanism B) produces a dispersive feature pinned to the anticrossing (dotted line). (b) Even versus odd drag magneto-oscillations for matched wires with the lumped small parameter $2\gamma\delta\eta_c=10^{-2}$, where $\delta$ is the saddle asymmetry and $\eta_c=\mathcal J_0/\mathcal J$ the coupling-asymmetry factor defined in Sec.~\ref{sec:nonreciprocal}; the odd/even ratio (dashed, right axis) grows with field as $\gamma\Delta\tau(B)\propto B/\Delta_1(B)$. (c) Drive-polarity asymmetry of the nonlinear drag, Eq.~\eqref{eq:NRnl}: dispersive zero crossings at the riser centers ($\bar T=\tfrac12$) and a plateau floor $\propto N_{\rm open}$, precisely where the reciprocal Fano factor (dotted) vanishes.}
\label{fig:nonrec}
\end{figure*}

The newest generation of experiments reports nonreciprocal drag~\cite{Makaju2024,Makaju2025,Cai2026}. The word is used for at least three physically distinct asymmetries, which our framework cleanly separates. 

(i) \emph{Drive-polarity asymmetry}, $I_D(V)\neq-I_D(-V)$: this is a nonlinear rectification effect, fully present for reciprocal transmissions [Eq.~\eqref{eq:nlin} is even in $V$], and it exists at zero magnetic field. It is the natural interpretation of the zero-field asymmetries of Refs.~\cite{Makaju2024,Zheng2025nonlinear}. 

(ii) \emph{Wire-exchange asymmetry}, $g_D^{12}\neq g_D^{21}$ at fixed $B$: Onsager reciprocity only constrains $g_D^{12}(B)=g_D^{21}(-B)$. Different confinements of the two wires do not by themselves produce an exchange asymmetry in linear response [Eq.~\eqref{eq:gdmain} is symmetric under $1\leftrightarrow2$ for any $\Delta_{1,2}^{(i)}$]; they enter only through the finite-bias mechanism discussed below.

(iii) \emph{Genuine transmission nonreciprocity}, $T_\to(\varepsilon)\neq T_\gets(\varepsilon)$ for the two transport directions of a single channel. As emphasized in Ref.~\cite{Solow2026}, the last is impossible for a two-terminal conductor with a unitary scattering matrix; it requires dissipation. Physically, the bath is the top or middle gate, or a dephasing environment~\cite{Buttiker1986,BrouwerBeenakker1997}. A gate held at fixed dc bias by a low-impedance source is an ac ground, and the dissipation is not dc particle exchange (the coupling is capacitive) but absorption of energy and coherence in the voltage-probe sense; phonons or charge two-level systems play the same role. For a channel uniformly coupled to such a featureless grounded bath at rate $\gamma$, the weak-dissipation result of Ref.~\cite{Solow2026} is
\begin{equation}
\Delta T(\varepsilon)\equiv T_\to-T_\gets
\simeq-2\gamma\,\bar T(\varepsilon)\,\Delta\tau(\varepsilon),
\label{eq:SF}
\end{equation}
where $\bar T=\tfrac12(T_\to+T_\gets)$ and $\Delta\tau=\tau_\to-\tau_\gets$ is the difference of directional transmission (Wigner) times~\cite{LandauerMartin1994}. Nonzero $\Delta\tau$ requires both time-reversal and inversion breaking (in practice, inversion breaking is supplied by an asymmetric gate-defined saddle or by disorder, while time reversal is broken by the applied field; disorder alone is not sufficient at $B=0$), and since the full system including the grounded bath is a reciprocal multiterminal conductor, $\Delta T$, and every drag effect built on it, is odd in magnetic field.

Before developing (iii), we pause on (ii) at zero field, where an exchange asymmetry $g_D^{12}\neq g_D^{21}$ has been reported explicitly~\cite{Makaju2024,Makaju2025,Zheng2026spin}. In strict linear response this asymmetry is forbidden: Onsager-Casimir reciprocity of the full system, wires, circuit, and any absorbing bath included as real terminals, gives $g_D^{12}(B)=g_D^{21}(-B)$, hence $g_D^{12}(0)=g_D^{21}(0)$ identically, and the formalism respects this manifestly [the product $A_1A_2$ and the kernel of Eq.~\eqref{eq:kernel} are exchange symmetric]. Dissipation cannot help at $B=0$, since all its effects are field odd. An observed zero-field exchange asymmetry is therefore itself evidence that the measurement is not in the linear regime, and at finite working bias the asymmetry appears naturally. What experiments extract is $V_{\rm drag}/I_{\rm drive}$ at finite drive. Expanding $I_D=g_DV+c_2V^2+O(V^3)$, the coefficient $c_2$ is the shot-noise term of Eq.~\eqref{eq:nlin}; it is built from the Fano factor of the \emph{drive} wire and from $\alpha_-$, both of which change under interchange of the roles of two nonidentical wires. The resulting fractional asymmetry, $|g^{12}_D-g^{21}_D|/g_D\sim|c_2^{(1)}-c_2^{(2)}|V/g_D\sim(eV\Delta_1/\pi T^2)\times O(\alpha_-/\alpha_+)$, is of order unity as soon as $eV$ exceeds the anomalously small crossover scale $eV^*\simeq\pi T^2/\Delta_1$, a condition met by typical nanoampere drive currents at dilution temperatures. A second, independent field-even mechanism is asymmetric heating: the Joule power is deposited primarily in the drive circuit, so the two wires sit at different effective temperatures, and which wire is heated swaps with the roles. Both mechanisms predict that the zero-field exchange asymmetry is a function of drive amplitude and must extrapolate to zero at vanishing bias, a directly testable statement.

Importing Eq.~\eqref{eq:SF} into the drag problem produces a qualitatively new rectification channel. Repeating the construction of Sec.~\ref{sec:dragformula} with direction-resolved transmissions, the rectification response of the drag wire splits into a dipole part, driven by the antisymmetric combination of the potential fluctuations on its two sides and controlled by the familiar asymmetry factor $A_2(\omega)$ built on $\bar T$, and a common-mode part,
\begin{equation}
A_{\Delta}(\omega)=\int d\varepsilon\,[f_--f_+]
\sum_n\big[\Delta T_n(\varepsilon_+)-\Delta T_n(\varepsilon_-)\big],
\label{eq:ADelta}
\end{equation}
driven by the symmetric combination: a dissipative nonreciprocal channel rectifies even fluctuations that push both of its sides equally, because its two counter-propagating rectified flows no longer cancel. A common shift of both reservoirs relative to the grounded bath is physical rather than pure gauge, and the channel disappears identically at $\gamma=0$. In the exact kernel language of Eq.~\eqref{eq:kernel}, the common-mode vertex couples with weight $\mathcal J_0=\sum_{ss'}s'|V_{ss'}|^2$, which vanishes for a mirror-symmetric interwire coupling: the effect requires a left/right asymmetry of the coupling geometry, quantified by $\eta_c=\mathcal J_0/\mathcal J$, in full accord with the Onsager constraint that a field-odd linear drag needs broken wire-exchange symmetry. The simplest realization is a longitudinal offset $x_0$ of the two bottlenecks, for which Appendix~\ref{app:kernel} gives $\eta_c\simeq-\sgn(x_0)(4\pi^2/5)(x_0/L_T)^2$: the coupling asymmetry is odd in the offset, as mirror symmetry requires, and reaches order unity for offsets of a fraction of the thermal length.

For an energy-smooth $\Delta\tau$ (the band-structure mechanism of Ref.~\cite{Solow2026}, in our setting a left/right asymmetry of the saddle), Eq.~\eqref{eq:ADelta} collapses onto the ordinary asymmetry factor exactly, $A_\Delta=-2\gamma\Delta\tau(B)\,A(\omega)$, and the field-odd fraction of the linear drag obeys a parameter-free relation,
\begin{equation}
\frac{\delta g_D^{\rm odd}}{g_D}
=\eta_c\left[\left(\frac{\Delta G}{G}\right)_{\!1}
+\left(\frac{\Delta G}{G}\right)_{\!2}\right],
\label{eq:flagship}
\end{equation}
where $(\Delta G/G)_i=-2\gamma_i\Delta\tau_i$ is the conductance nonreciprocity of wire $i$ measured separately. The field-odd drag oscillates in phase with $g_D(h)$, with an envelope that grows with field since the traversal time scales with the inverse tunneling width, $\Delta\tau\sim\delta/\Delta_1(B)$ [Fig.~\ref{fig:nonrec}(b)]. If instead $\Delta\tau(\varepsilon)$ is resonant, the interference mechanism of Ref.~\cite{Solow2026}, realized here at anticrossings of the magnetoelectric subbands, the odd component acquires a dispersive lineshape pinned to the anticrossing [Fig.~\ref{fig:nonrec}(a)]: the lineshape of the field-odd drag diagnoses the microscopic origin of the transmission-time difference.

In the nonlinear regime the drive polarity selects the transport direction in the drive wire, so Eq.~\eqref{eq:nlin} becomes direction-resolved and the polarity asymmetry at fixed $|V|$ reads
\begin{equation}
I_D(V)-I_D(-V)=-2\gamma\Delta\tau\,
\frac{eV^2\alpha_-(0)}{\Delta_2^{\rm c}R_Q}
\sum_n\bar T_n(1-2\bar T_n).
\label{eq:NRnl}
\end{equation}
Its gate dependence [Fig.~\ref{fig:nonrec}(c)] is distinctive: it crosses zero at each conductance step ($\bar T=\tfrac12$) and tends to a floor proportional to the number of open channels on the conductance plateaus -- exactly where the reciprocal nonlinear drag $\propto\sum\bar T(1-\bar T)$ vanishes. The nonreciprocal polarity asymmetry is therefore best sought on the plateaus, where the reciprocal background disappears. On the plateaus a quantitative treatment requires the noise theory of absorbing conductors~\cite{BeenakkerBrouwer2001}; the conductance step-vicinity structure of Eq.~\eqref{eq:NRnl} is the controlled part. Crucially, Eq.~\eqref{eq:NRnl} is odd in $B$, whereas the $\alpha_-$ polarity asymmetry of the reciprocal theory is even in $B$: antisymmetrizing the measured polarity asymmetry in field cleanly separates the two contributions. This is, in our view, the sharpest experimental test available to the ballistic-wire experiment of Ref.~\cite{Cai2026}.

Finally, consider the zero-field limit. Strictly at $B=0$ with time-reversal symmetry, the scattering matrix is symmetric and the nonreciprocity $\Delta T$ of Eq.~\eqref{eq:SF} vanishes identically even at finite $\gamma$: all field-odd channels are absent, and any zero-field nonreciprocity of drag is the reciprocal nonlinear effect (i). The interesting zero-field statement concerns the onset of the field-odd component as the field is turned on: for wires with Rashba spin--orbit coupling, an in-plane field skewed from the spin--orbit axis activates $\Delta\tau\propto B_\parallel$ linearly~\cite{Solow2026}, predicting a linear-in-$B$ magnetochiral drag near $B=0$ with slope proportional to the gate-tunable dissipation $\gamma$, against the quadratic orbital background of Sec.~\ref{sec:oscillations}, a smoking gun of the non-Hermitian channel. The conductance nonreciprocity has been estimated at $10^{-3}$--$10^{-4}\,G_Q$ for realistic parameters (InAs nanowires with $v_F\sim10^5\,$m/s, micron lengths, Rashba--Zeeman anticrossings, and weak uniform dissipation $\gamma\sim\mu$eV)~\cite{Solow2026}. Equation~\eqref{eq:flagship} then puts the field-odd drag fraction at $\eta_c\times10^{-3}$, at the edge of current sensitivity but greatly aided by the background-free plateau regions of Eq.~\eqref{eq:NRnl}.

%###################################################################################
%###################################################################################
%###################################################################################
%###################################################################################
%###################################################################################
\section{Further extensions}\label{sec:extensions}

\subsection{Zeeman splitting}\label{sec:zeeman}

\begin{figure*}[t]
\begin{minipage}{0.32\textwidth}\textbf{(a)}\\ \includegraphics[width=\linewidth]{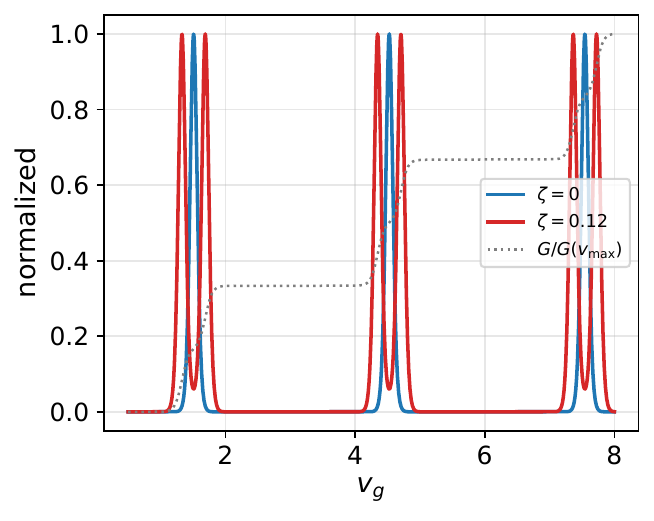}\end{minipage}\hfill
\begin{minipage}{0.32\textwidth}\textbf{(b)}\\ \includegraphics[width=\linewidth]{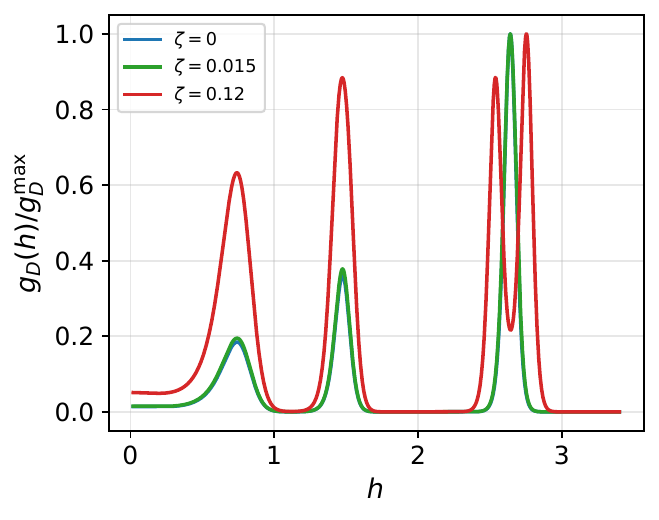}\end{minipage}\hfill
\begin{minipage}{0.32\textwidth}\textbf{(c)}\\ \includegraphics[width=\linewidth]{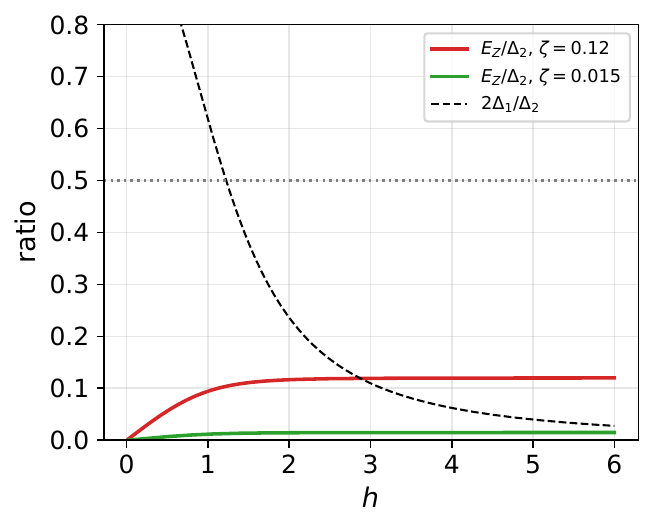}\end{minipage}
\caption{Zeeman splitting of the drag oscillations, Eq.~\eqref{eq:zeeman}. (a) Gate sweeps at $h=3$: each drag peak splits into a doublet for the exchange-enhanced $\zeta=0.12$; the conductance staircase (dotted) develops steps of $\tfrac12G_Q$. (b) Field sweeps at $v_g=4$: the highest-field peak develops a resolved doublet for $\zeta=0.12$ but not for the bare GaAs value $\zeta=0.015$. (c) The governing ratios: the splitting-to-spacing ratio $E_Z/\Delta_2$ saturates at $\zeta$, while the riser width-to-spacing ratio $2\Delta_1/\Delta_2\simeq a/h^2$ falls with field; doublets resolve where the curves cross.}
\label{fig:zeeman}
\end{figure*}

Spin enters Eqs.~\eqref{eq:gdmain} and \eqref{eq:transcond} through spin-resolved subband bottoms, each spin channel carrying half weight:
\begin{equation}
\Delta_2(B)(n+\tfrac12)\;\longrightarrow\;
\Delta_2(B)(n+\tfrac12)\pm\tfrac12 E_Z,
\quad
E_Z=\zeta h\,\omega_x,
\label{eq:zeeman}
\end{equation}
with $\zeta=g^*m^*/2m_e\simeq0.015$ for bare GaAs and up to $\zeta\sim0.1$--$0.2$ with exchange enhancement~\cite{Thomas1996}. The structure of the effect is controlled by two ratios [Fig.~\ref{fig:zeeman}(c)]. The splitting-to-spacing ratio $E_Z/\Delta_2\to\zeta$ saturates at large field, since both scale linearly with $B$: Zeeman effect never reorders the spectrum. Resolvability is instead governed by the conductance step width, $2\Delta_1/\Delta_2\simeq a/h^2$, which falls with field; doublets therefore emerge above a threshold field $h^*$ determined by $\zeta h^*\simeq2\Delta_1(h^*)$, the crossing of the two curves in Fig.~\ref{fig:zeeman}(c), i.e., $h^*\simeq\sqrt{a/\zeta}$. Numerically $h^*\approx2.9$ for $\zeta=0.12$ but $h^*\approx8.2$ for the bare $g^*$ ($a=1$): resolved spin doublets in the drag oscillations at accessible fields are themselves evidence of an exchange-enhanced $g$ factor, making drag a spin spectrometer for the constriction. Above $h^*$ every drag peak (in gate or field sweeps, linear or nonlinear) splits into a symmetric doublet (Fig.~\ref{fig:zeeman}). Since a perpendicular field alone cannot reach the coincidence condition $E_Z=\Delta_2/2$ (it would require $\zeta=\tfrac12$), tilting the field provides an independent Zeeman knob; at coincidence the drag oscillation period in gate voltage halves, a drag analog of the Shubnikov--de Haas coincidence method. These considerations connect directly to the spin-resolved drag features reported in Ref.~\cite{Zheng2026spin}.

\subsection{Interaction renormalization of the constriction transmission}\label{sec:myg}

\begin{figure*}[t]
\begin{minipage}{0.44\textwidth}\textbf{(a)}\\ \includegraphics[width=\linewidth]{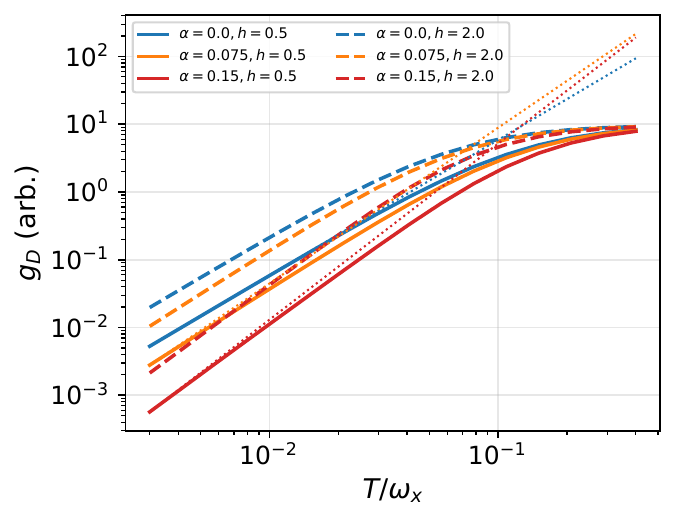}\end{minipage}\hfill
\begin{minipage}{0.44\textwidth}\textbf{(b)}\\ \includegraphics[width=\linewidth]{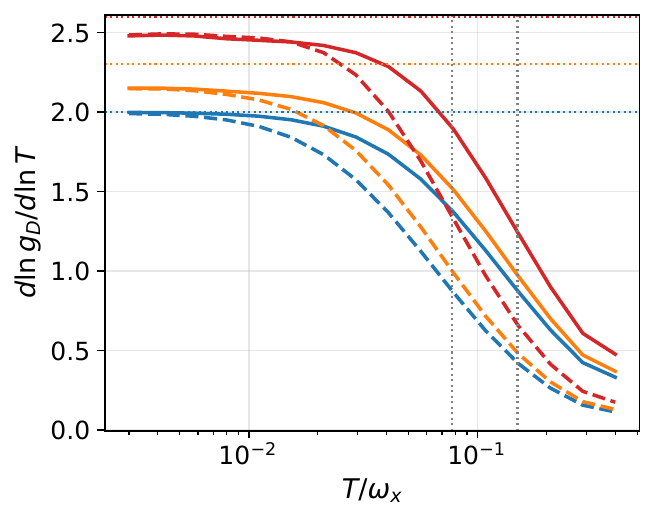}\end{minipage}
\caption{Luttinger renormalization combined with magnetic field. (a) Drag $g_D(T)$ at the riser for interaction strengths $\alpha=0,\,0.075,\,0.15$ and two fields; the dotted guides show $T^{2+4\alpha}$. (b) The local exponent $d\ln g_D/d\ln T$: anomalous plateaus at low $T$ approach $2+4\alpha$ (horizontal dotted lines), and the crossover scale $\Delta_1(B)/\pi$ (vertical lines) moves down with field, at fixed $T$, increasing $B$ lowers the apparent drag exponent.}
\label{fig:myg}
\end{figure*}

Electron interactions within each wire renormalize the barrier transmission. For a weakly reflecting barrier embedded in a one-dimensional channel with a short-range interaction of dimensionless strength $\alpha$, the treatment of Refs.~\cite{Matveev1993,Yue1994,KaneFisher1992} yields the renormalization-group--improved transmission
\begin{equation}
T_R(\varepsilon;T)=\frac{T_0(\varepsilon)\,\Lambda^{2\alpha}}
{1-T_0(\varepsilon)+T_0(\varepsilon)\,\Lambda^{2\alpha}},
\;\;
\Lambda=\frac{\max(|\varepsilon\!-\!\mu|,T)}{D},
\label{eq:myg}
\end{equation}
with $T_0$ the bare ($B$-dependent) saddle transmission and $D$ a bandwidth cutoff, of order of the smaller of the one-dimensional subband depth ($E_F$ measured from the subband bottom, typically 1--5\,meV) and $v_F/\ell_c$; the numerics below use $D=20\,\omega_x$. A word on applicability is in order, since Refs.~\cite{Matveev1993,Yue1994} treat a pointlike scatterer in a long uniform channel while our barrier is the constriction itself. The resolution lies in the separation of scales built into the devices. The constriction created by local gating has a length $\ell_c$ short compared with the wire length $L$. At energies below $v_F/\ell_c$ the entire constriction therefore acts as a single effective scatterer embedded in a long interacting channel, and Eq.~\eqref{eq:myg} applies with the constriction playing the role of the barrier. In this window the model is qualitatively, and for weak interaction quantitatively, controlled; at energies above $v_F/\ell_c$ the internal structure of the constriction resolves and the renormalization saturates, which only cuts off the flow at $\Lambda\sim v_F/\ell_cD$ without changing the low-energy behavior. The exponent is tied to the Luttinger parameters of a single wire: for weak coupling $\alpha=[U(0)-U(2k_F)]/2\pi v_F\simeq1-K$ (spinless), and for spinful wires $\alpha\to(K_c^{-1}+K_s^{-1}-2)/2\simeq(2-K_c-K_s)/2$ in terms of the charge and spin parameters, i.e., $2\alpha\simeq1-K_c$ for spin-rotation-invariant wires with $K_s=1$. It is distinct from the interwire parameters $K_c^\pm$ of the coupled pair extracted in Refs.~\cite{Laroche2014,Makaju2025}, which govern the momentum-transfer channel of Sec.~\ref{sec:edge}. Parametrically, $\alpha$ depends on gate voltage and field through $v_F$ and the screened interaction, both of which enhance $\alpha$ (reduce $K$) at low channel density.

Feeding Eq.~\eqref{eq:myg} into the exact asymmetry factor Eq.~\eqref{eq:Afactor} and the frequency integral Eq.~\eqref{eq:master} yields the drag at the conductance step without further approximation (Fig.~\ref{fig:myg}). Each wire's asymmetry factor at resonance scales as $A\propto T^{1+2\alpha}$, hence
\begin{equation}
g_D^{\rm peak}(T)\propto T^{\,2+4\alpha},
\qquad \pi T\ll\Delta_1(B),
\label{eq:anomexp}
\end{equation}
crossing over to the smeared regime above $\Delta_1(B)/\pi$. The numerics confirm the anomalous plateaus (local exponents $2.0$, $\simeq2.3$, $\simeq2.5$ for $\alpha=0,\,0.075,\,0.15$, approaching $2+4\alpha$ with slow logarithmic corrections). The new observation is the field-tuned crossover: since $\Delta_1(B)/\pi$ decreases with field, a field sweep at fixed temperature carries the system out of the anomalous regime, the measured drag exponent becomes a function of $B$. The anomalous power is not confined to low temperatures: in the smeared regime $\pi T\gtrsim\Delta_1(B)$, where the peak drag is $T$ independent for noninteracting wires [Eq.~\eqref{eq:gdhighT}], the renormalization cut at $T$ retains the factor $[T_R(\mu;T)]^2$ per wire and gives a slow anomalous rise $g_D^{\rm peak}\propto T^{4\alpha}$; the effective exponent therefore interpolates continuously from $2+4\alpha$ down toward $4\alpha$ [visible in Fig.~\ref{fig:myg}(b)]. The magnetic field thus provides a gate-free experimental knob on the Luttinger-liquid drag exponents, and, combined with the backscattering channel of Sec.~\ref{sec:edge}, a natural framework for the anomalous, nonmonotonic temperature dependences reported in Refs.~\cite{Laroche2014,Zheng2025nonlinear}.

\subsection{High-field regime: backscattering versus rectification}\label{sec:edge}

\begin{figure*}[t]
\begin{minipage}{0.44\textwidth}\textbf{(a)}\\ \includegraphics[width=\linewidth]{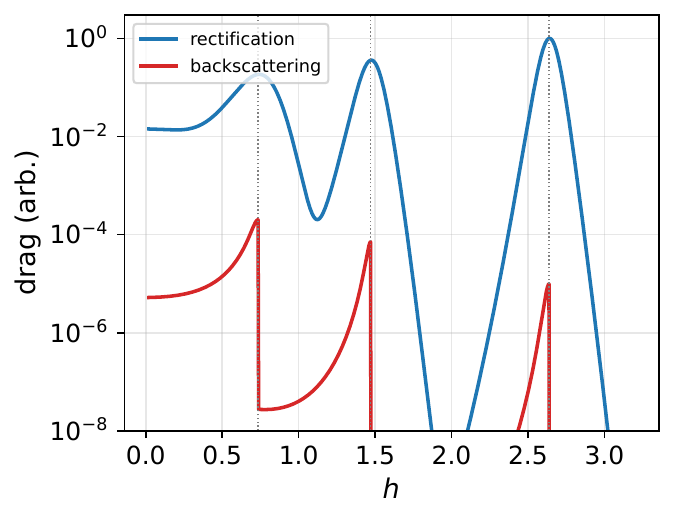}\end{minipage}\hfill
\begin{minipage}{0.44\textwidth}\textbf{(b)}\\ \includegraphics[width=\linewidth]{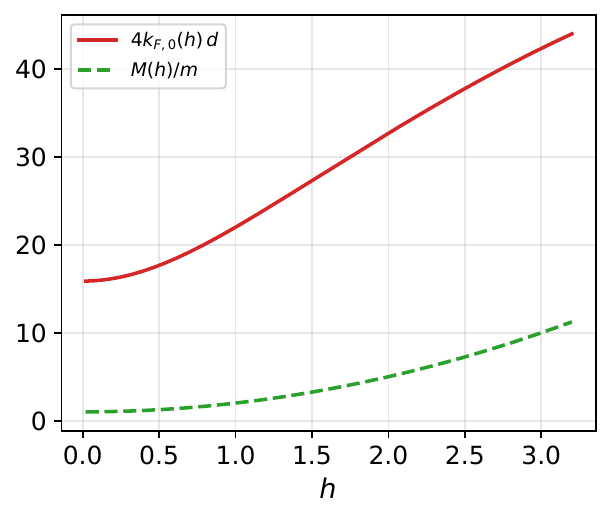}\end{minipage}
\caption{Backscattering versus rectification drag at high field ($v_g=4$, $d/\ell_x=1.5$, $T/\omega_x=0.05$, $a=1$). (a) The rectification drag (symmetric peaks) and the momentum-transfer estimate Eq.~\eqref{eq:bs} (one-sided sawtooth peaks) on a logarithmic scale; dotted verticals mark the depopulation fields. (b) The suppression exponent $4k_{F,0}(h)d$ and the mass enhancement $M(h)/m$ grow with field, so the rectification mechanism increasingly dominates.}
\label{fig:edge}
\end{figure*}

At $\omega_c\gtrsim\omega_{x,y}$ the constriction carries magnetoelectric (edge-like) channels with dispersion $E_n(q)=\Delta_2(h)(n+\tfrac12)+q^2/2M(h)$, $q$ being the momentum along the channel, and the field-enhanced one-dimensional mass $M(h)/m=(h^2+a^2)/a^2$. In this regime the momentum-transfer mechanism of drag between the wires via interwire backscattering with momentum transfer $2k_F$~\cite{KlesseStern2000,Dmitriev2012} competes with rectification.  For wires at distance $d$ the relevant matrix element carries $|U(2k_F)|^2\propto e^{-4k_Fd}$, giving the weak-coupling estimate per channel
\begin{gather}
g_D^{\rm bs}\;\propto\;u^2\,T\,e^{-4k_{F,n}(h)\,d},
\nonumber\\
k_{F,n}(h)=\sqrt{2M(h)\,\big[\mu-\Delta_2(h)(n+\tfrac12)\big]},
\label{eq:bs}
\end{gather}
Here $\mu$ is the global Fermi level set by the density of the two-dimensional reservoirs, so the carrier density in the channel, and with it $k_{F,n}$, is fixed jointly by the 2DEG density (through $\mu$) and by the gates (through $V_g$ and the screening that determines $\omega_{x,y}$). Two consequences follow (Fig.~\ref{fig:edge}). First, the exponential suppression is lifted only as a channel approaches depopulation ($k_F\to0$): the backscattering drag peaks at the same fields as the rectification drag, but with a one-sided sawtooth lineshape (a rise on the occupied side terminated by the depopulation cutoff) sharply distinguishable from the symmetric rectification peaks. Second, away from the thresholds the suppression strengthens with field through $M(h)$, so rectification increasingly dominates the oscillation pattern as $B$ grows. The discrimination between the two mechanisms is therefore twofold: lineshape (sawtooth versus symmetric) and temperature dependence ($T$ versus $T^2$ at weak coupling; with Luttinger corrections, $T^{4K_c^--3}$, with $K_c^-$ the interwire relative-charge parameter~\cite{KlesseStern2000,Laroche2014} versus $T^{2+4\alpha}$ of Sec.~\ref{sec:myg}).

The two channels also resolve a question raised by the experiments: none of the rectification results of this paper produces a drag that grows on cooling, yet such an upturn is reported at the lowest temperatures~\cite{Laroche2014,Zheng2025nonlinear}. The upturn lives in the backscattering channel once Luttinger corrections are kept. The total drag is the sum of the two contributions,
\begin{equation}
g_D(T)\;\simeq\;a_{\rm rect}\,T^{\,2+4\alpha}
\;+\;a_{\rm bs}\,e^{-4k_Fd}\;T^{4K_c^--3},
\label{eq:composite}
\end{equation}
and the second term diverges on cooling for $K_c^-<3/4$: the total is generically nonmonotonic, with a minimum at the crossover temperature $T_*$ obtained by equating the two terms, below which the drag turns up. The position of the minimum is predicted to move with magnetic field through both $\Delta_1(B)$ (which controls $a_{\rm rect}$) and $k_F(B)$ (which controls the exponential); a quantitative confrontation requires the spinful two-sector exponents and is left for future work. The distance dependences of the two terms are equally discriminating: the rectification drag depends on the wire separation only through the zero-momentum coupling, $g_D^{\rm rect}\propto\hat{\mathcal V}^2(0)\propto\ln^2(\lambda_s/d)$ for gate-screening length $\lambda_s>d$, i.e., logarithmically, while the backscattering term falls exponentially, $e^{-4k_Fd}$. Comparing otherwise similar lateral devices with different separations therefore separates the channels directly.

\subsection{Frequency structure of the circuit coupling}\label{sec:circuit}

\begin{figure*}[t]
\begin{minipage}{0.44\textwidth}\textbf{(a)}\\ \includegraphics[width=\linewidth]{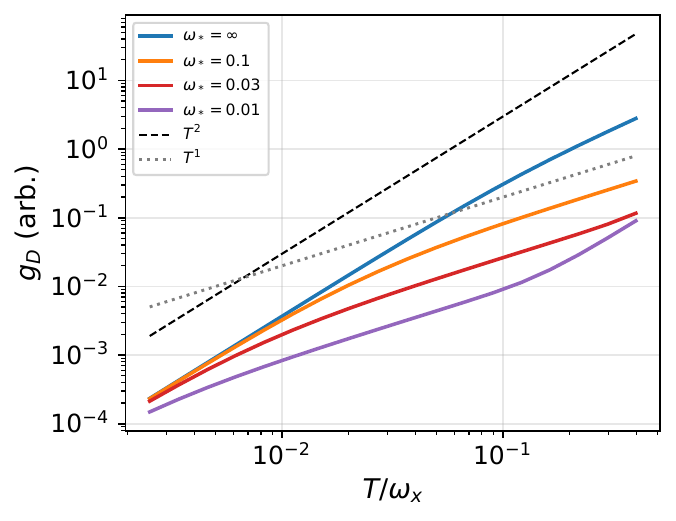}\end{minipage}\hfill
\begin{minipage}{0.44\textwidth}\textbf{(b)}\\ \includegraphics[width=\linewidth]{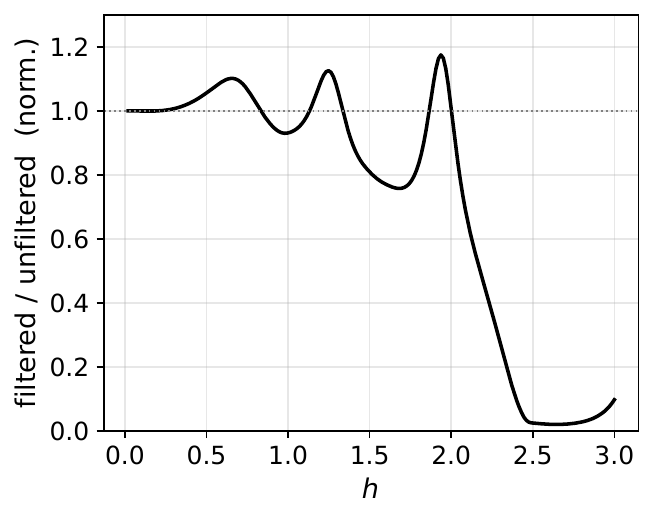}\end{minipage}
\caption{Effect of the frequency structure of the circuit kernel. (a) Drag $g_D(T)$ at the riser for several $RC$ cutoffs $\omega_*$ (in units of $\omega_x$): the $T^2$ law bends over to $T$ at $T\sim\omega_*$. (b) Ratio of the normalized magneto-oscillation traces with and without filtering ($T/\omega_x=0.04$, $v_{g,1}=v_{g,2}=5$): the ratio stays near unity -- the oscillation shape is unaffected by the circuit filter -- and deviates only at high field where $\Delta_1(B)$ approaches $T$.}
\label{fig:circuit}
\end{figure*}

The results above assume that the interaction kernel is frequency independent over the thermal window $\omega\sim T$. In a real circuit the coupling is dressed by trans-impedances with $RC$ structure, $\alpha_+(\omega)=\alpha_+(0)/[1+(\omega/\omega_*)^2]^2$, where $\omega_*=1/RC$ is set by the relevant mutual and self-capacitances ($\sim$aF--fF) and the impedance of the leads and environment ($\sim$k$\Omega$--M$\Omega$). For M$\Omega$-range environments this gives $\omega_*\sim1$--$10^3\,\mu$eV, comparable to $T$ at dilution temperatures; for k$\Omega$ leads $\omega_*\gg T$ and the filter is inactive. Evaluating the full frequency integral with the exact rectification coefficients we find two results (Fig.~\ref{fig:circuit}). First, once the circuit cutoff drops below the thermal frequency, $\omega_*<T$, the drag integral is cut at $\omega_*$ rather than $T$ and the linear-regime law softens,
\begin{equation}
g_D\propto\frac1T\int^{\omega_*}\!\!d\omega\,\omega^2\Big(\frac{2T}{\omega}\Big)^{\!2}\propto T\,\omega_*\,,
\end{equation}
a $T^2\!\to\!T$ conversion with no interaction physics involved. Second, the filter factorizes out of the magneto-oscillations as long as $\Delta_1(B)\gg T$: the normalized $g_D(h)/g_D(0)$ traces with and without filtering coincide, with deviations confined to high fields where $\Delta_1(B)\sim T$. Together with Secs.~\ref{sec:myg} and \ref{sec:edge} this completes a three-way diagnostic for any anomalous drag temperature law: circuit filtering gives $T$ with intact oscillation shapes; interaction renormalization gives $T^{2+4\alpha}$ with intact shapes but a field-tunable exponent; backscattering modifies the lineshapes themselves.

%###################################################################################
%###################################################################################
%###################################################################################
%###################################################################################
%###################################################################################
\section{Summary, conclusions, and outlook}\label{sec:summary}

We have developed a unified theory of Coulomb drag between interactively coupled adiabatic constrictions in a magnetic field, spanning the linear and nonlinear transport regimes and reciprocal as well as nonreciprocal responses. The physical picture that organizes all of the results is mesoscopic rectification: each circuit converts the electric fluctuations of the other into a dc current with an efficiency set by the particle-hole asymmetry of its transmission, which is sharply peaked where a magnetoelectric subband crosses the Fermi level. From this single principle follow: (i) drag magneto-oscillations locked to the magnetic-depopulation fields, asymptotically periodic in $1/B$, with contrast controlled by the ratio of temperature to the field-sharpened tunneling width $\Delta_1(B)$, and described quantitatively by the closed-form Eq.~\eqref{eq:gdmain}; (ii) a nonlinear regime, entered at bias voltages parametrically smaller than temperature, in which the drag current is a map of the drive channel's transconductance at any field, Eq.~\eqref{eq:transcond}, with oscillation envelopes that grow with field in contrast to the linear regime; (iii) a family of nonreciprocal drag effects enabled by dissipation, odd in magnetic field, obeying the parameter-free relation Eq.~\eqref{eq:flagship} between the drag nonreciprocity and the separately measurable conductance nonreciprocities of the individual wires; and (iv) sharp diagnostics (Zeeman doublet thresholds, field-tunable anomalous exponents, sawtooth versus symmetric lineshapes, and $T$ circuit conversion) that discriminate among the mechanisms that can underlie anomalous temperature dependences of drag.

Although these results are derived within a particular model, the exactly solvable saddle-point constriction, several of the conclusions are, we believe, generic. The oscillation principle requires only that transmission steps sweep through the Fermi level as field or gates vary; any quasi-one-dimensional two-subsystem device with quantized conductance will exhibit the same drag combs, with Eq.~\eqref{eq:gdmain} as the universal lineshape. The transconductance identity Eq.~\eqref{eq:transcond} relies only on the two-terminal noise formula and the Fermi-function shape of a thermally or curvature-broadened transmission step, both robust beyond the saddle model. The nonreciprocity relations Eqs.~\eqref{eq:flagship} and \eqref{eq:NRnl} follow from symmetry (Onsager reciprocity of the bath-inclusive system) plus the weak-dissipation form of $\Delta T$, and their field-parity signatures (the sharpest experimental discriminators we propose) are entirely model independent. Likewise the sign of the drag: the exact kernel Eq.~\eqref{eq:kernel} is not positive definite, being a difference of same-side and cross-side coherences, so the theory naturally accommodates the negative drag reported in several experiments~\cite{Yamamoto2006,Laroche2011} without invoking exotic correlations, the sign is a geometric property of the interwire coupling.

The most immediate points of contact are with the drag experiments on gate-defined double quantum wires reported in Refs.~\cite{Laroche2011,Laroche2014,Makaju2024,Zheng2025nonlinear,Makaju2025,Zheng2026spin,Cai2026}. The measured drag-versus-gate-and-field maps can be compared directly with Fig.~\ref{fig:params}(c) and with the transconductance identity of Sec.~\ref{sec:nonlinear}; the nonmonotonic temperature dependences invite the field-tunable-exponent analysis of Sec.~\ref{sec:myg}; the spin-resolved features connect to the Zeeman thresholds of Sec.~\ref{sec:zeeman}; and the reported nonreciprocities can be decomposed, by field antisymmetrization and polarity analysis, into the reciprocal-nonlinear and dissipation-induced components classified in Sec.~\ref{sec:nonreciprocal}. On the theory side, three open problems emerge from the confrontation with the data. (i) The zero-field exchange asymmetry $g_D^{12}\neq g_D^{21}$: forbidden in strict linear response (Sec.~\ref{sec:nonreciprocal}), it calls for a finite-bias theory of the coupled wires with self-consistent heating, whose sharpest prediction (the vanishing of the asymmetry under zero-bias extrapolation) is immediately testable. (ii) The nonmonotonic temperature dependence and the drag upturn at the lowest temperatures: Eq.~\eqref{eq:composite} provides the skeleton, but a quantitative theory requires the spinful two-sector Luttinger exponents of the coupled pair together with the field dependence of both channels. (iii) The nonreciprocal sector: a microscopic treatment of the absorbing channel in the scattering-state formalism, and the noise theory of the absorbing constriction on the conductance plateaus, would replace the weak-dissipation phenomenology of Sec.~\ref{sec:nonreciprocal} by a controlled calculation and extend it to the strongly interacting regime.

\begin{acknowledgments}
The work of A. L. was supported by NSF Grant No. DMR-2452658 and H. I. Romnes Faculty Fellowship provided by the University of Wisconsin-Madison Office of the Vice Chancellor for Research and Graduate Education with funding from the Wisconsin Alumni Research Foundation.  D. L. and M. Z were supported by the National Science Foundation through NSF/DMR-2518016. This work was performed in part during the workshop program "Emerging New Phases in Quantum Materials: The Disordered, the Strange and the Topological" at the Aspen Center for Physics, which is supported by National Science Foundation grant PHY-2210452. Portions of this work were carried out with the assistance of the large language model Claude (Anthropic) used to perform supporting numerical calculations and prepare figures, and to assist in editing the manuscript. The authors conceived the project and carried out the research.  

\end{acknowledgments}

%###################################################################################
%###################################################################################
%###################################################################################
%###################################################################################
%###################################################################################

\appendix

\section{Linear-response formalism}\label{app:formalism}

This appendix summarizes the derivation of the central formula Eq.~\eqref{eq:master}; details follow the Keldysh scattering-state construction of the quantum-circuit drag problem~\cite{Levchenko2008prl,Kamenev1995,Flensberg1995}.

Each constriction is described by its exact scattering states. For a channel with transmission and reflection amplitudes $t_k$, $r_k$, the field operator is expanded in the states $u_k(x)$, $v_k(x)$ incident from the left and right, and the current and density operators are projected onto the slowly varying (non-$2k_F$) components; the rapidly oscillating Friedel terms, which carry factors $e^{\pm2ik_Fx}$, average out after the spatial integrations against an interaction smooth on the Fermi wavelength and are dropped throughout (their physics, interwire backscattering, is estimated separately in Sec.~\ref{sec:edge}). The dynamics is encoded in the Keldysh partition function with two auxiliary fields coupling to the density and current of each wire; expanding to first order in the current vertex of the drag wire and to second order in the interwire interaction, and keeping the voltage-linear part of the drive-wire polarization operator, one arrives at
\begin{align}
g_D&=\frac{1}{8\pi T}\int_{-\infty}^{+\infty}
\frac{d\omega}{\sinh^2(\omega/2T)}
\iiiint dx_1dx_2dx_3dx_4
\nonumber\\
&\times\Gamma^K_2(x_1,x_2,\omega)\,\Gamma^K_1(x_3,x_4,\omega)\,
\mathcal V(x_1,x_3)\,\mathcal V(x_2,x_4),
\label{eq:masterapp}
\end{align}
where the interaction is treated as static (retardation is addressed in Sec.~\ref{sec:circuit}) and the triangular (rectification) vertex of wire $i$, evaluated on the smooth components, is
\begin{align}
\Gamma^K_i(x,x',\omega)=-\frac{e\,A_i(\omega)}{8\pi v_F^2}
\Big[&\theta(x)\theta(x')\,e^{i\omega(x-x')/v_F}
\nonumber\\
-\,&\theta(-x)\theta(-x')\,e^{-i\omega(x-x')/v_F}\Big],
\label{eq:vertexapp}
\end{align}
with the particle-hole asymmetry factor $A_i(\omega)$ of Eq.~\eqref{eq:Afactor}. A bias $V$ applied to wire 1 enters through its distribution function, $f\to\tfrac12[f(\varepsilon-eV/2)+f(\varepsilon+eV/2)]$, inside the wire-1 polarization: the $V$-linear term reproduces Eq.~\eqref{eq:masterapp}, while the $V^2$ term is the shot-noise contribution of Sec.~\ref{sec:nonlinear} -- this is also the entry point for the role-asymmetric effects discussed in Sec.~\ref{sec:nonreciprocal}. The two-sided structure of Eq.~\eqref{eq:vertexapp}, equal magnitudes and opposite phases on the two sides of the barrier, embodies current conservation: the rectified charge is emitted symmetrically into the two leads with opposite sign of the induced current. For the saddle-point transmission Eq.~\eqref{eq:kemble}, which has the form of a Fermi--Dirac step of width $\Delta_1/\pi$ as a function of energy, the frequency dependence of $A_i$ is known in closed form,
\begin{gather}
A_i(\omega)=\bar\Delta_i\sum_n
\ln\!\left[1+\frac{\sinh^2(\omega/2\bar\Delta_i)}
{\cosh^2 w_{in}}\right],
\nonumber\\
w_{in}=\big(v_{g,i}\omega_x-\Delta_2^{(i)}(n+\tfrac12)\big)/2\bar\Delta_i,
\label{eq:Aclosed}
\end{gather}
with $\bar\Delta_i=\max(\Delta_1^{(i)}/\pi,\,T)$ interpolating between the curvature-broadened ($\pi T\ll\Delta_1$) and thermally broadened ($\pi T\gg\Delta_1$) limits; as defined in Eq.~\eqref{eq:Afactor}, $A_i(\omega)$ is non-negative and even in $\omega$. At small frequencies
\begin{equation}
A_i\simeq\frac{\omega^2}{4\bar\Delta_i}\sum_n\cosh^{-2}\big[w_{in}\big],
\end{equation}
with the same argument $w_{in}=(v_{g,i}\omega_x-\Delta_2^{(i)}(n+1/2))/2\bar{\Delta}_i$ as in Eq.~\eqref{eq:Aclosed} -- the distance of the Fermi level of wire $i$ from the $n$th riser in units of twice the step width. This asymptote produces the $\cosh^{-2}$ combs of Eq.~\eqref{eq:gdmain} after the frequency integration $\int dz\,z^2\,\sinh^{-2}(z/2)=8\pi^2/3$.

\section{Exact spatial kernel and the wire limit}\label{app:kernel}

\begin{figure*}[t!]
\begin{minipage}{0.32\textwidth}\textbf{(a)}\\ \includegraphics[width=\linewidth]{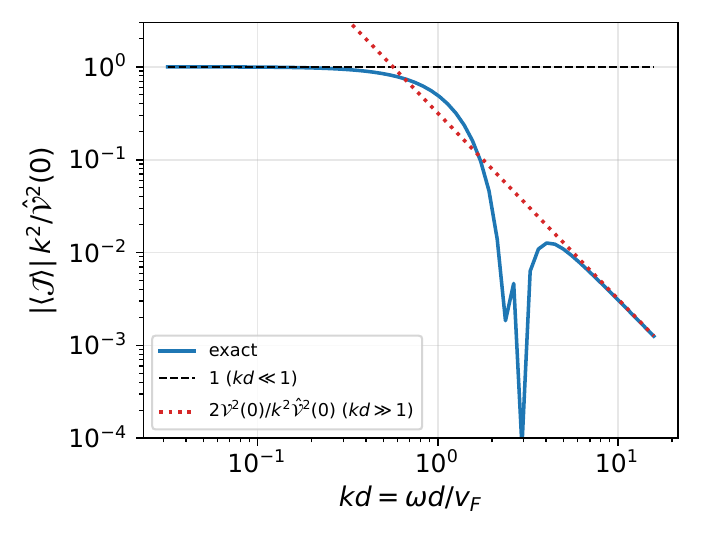}\end{minipage}\hfill
\begin{minipage}{0.32\textwidth}\textbf{(b)}\\ \includegraphics[width=\linewidth]{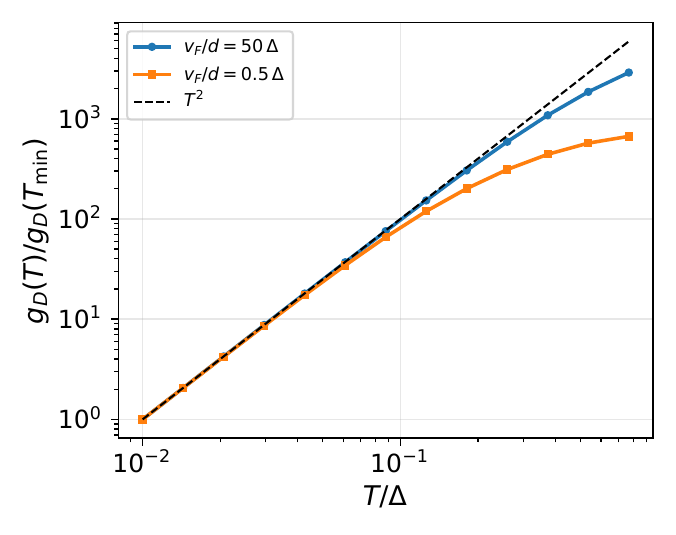}\end{minipage}\hfill
\begin{minipage}{0.32\textwidth}\textbf{(c)}\\ \includegraphics[width=\linewidth]{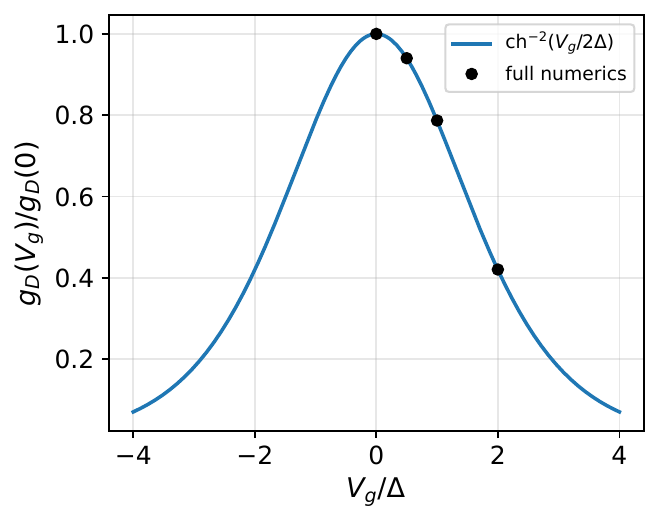}\end{minipage}
\caption{Numerical validation of the exact kernel. (a) The phase-averaged wire kernel $\langle\mathcal J\rangle k^2/\hat{\mathcal V}(0)^2$ versus $kd$ for a Gaussian interaction: plateau at unity for $kd\ll1$ [Eq.~\eqref{eq:Jwire}], crossover to the $2\mathcal V(0)^2/k^2\hat{\mathcal V}(0)^2$ tail for $kd\gg1$; the dip marks a sign change of $\mathcal J$. (b) Full $g_D(T)$ from Eqs.~\eqref{eq:masterapp} and \eqref{eq:Jwire} with the exact $A(\omega,T)$, normalized to its value at $T_{\min}=10^{-2}\Delta$ and shown for $T<\Delta$: the $T^2$ law (dashed guide) holds for $T\ll\min(\Delta,v_F/d)/\pi$ and softens beyond. (c) Gate dependence of the full numerical evaluation (points) against the $\cosh^{-2}(V_g/2\Delta)$ lineshape (line), with $\Delta$ the transmission-step width of the validation model.}
\label{fig:kernel}
\end{figure*}

Writing Eq.~\eqref{eq:vertexapp} as $\Gamma^K_i=-c\,A_i\sum_{s=\pm}s\,\theta_s(x)\theta_s(x')e^{isk(x-x')}$ with $\theta_\pm(x)=\theta(\pm x)$, $c=e/8\pi v_F^2$, and $k=\omega/v_F$, the four spatial integrals in Eq.~\eqref{eq:masterapp} factorize into the pairs $(x_1,x_3)$ and $(x_2,x_4)$ connected by the two interaction legs:
\begin{equation}
\iiiint\Gamma^K_2\Gamma^K_1\,\mathcal V\mathcal V
=c^2A_1A_2\sum_{s,s'}s\,s'\,|V_{ss'}(k)|^2,
\label{eq:factorization}
\end{equation}
with $V_{ss'}(k)$ defined in Eq.~\eqref{eq:kernel}. Equation~\eqref{eq:factorization} is an identity: no assumption about the range, shape, or symmetry of $\mathcal V(x,y)$ is involved. Three structural properties follow immediately. (i) \emph{Gauge invariance}: for $\mathcal V=\mathrm{const}$ all four $V_{ss'}$ coincide and $\mathcal J\equiv0$ -- a uniform potential drives no current. (ii) \emph{Circuit correspondence}: for lead-resolved coupling, $\mathcal V(x,y)=\mathcal V_{ab}$ with $x$ on side $a$ and $y$ on side $b$, the kernel reduces to $|\hat g(k)|^4[\mathcal V_{LL}^2+\mathcal V_{RR}^2-2\mathcal V_{LR}\mathcal V_{RL}]$. This is precisely the trans-impedance combination entering $\alpha_+$ of Ref.~\cite{Levchenko2008prl}. (iii) \emph{No positivity}: $\mathcal J$ is a difference of same-side and cross-side coherences and changes sign as a function of $k$ and of the coupling geometry; since the remaining factors in Eq.~\eqref{eq:master} are positive, the sign of the linear drag is not fixed by any fundamental principle within this mechanism. The model therefore accommodates negative drag~\cite{Yamamoto2006,Laroche2011} as a geometric property of the interwire coupling, without invoking additional correlation physics.

For the translationally invariant coupling of the double-wire geometry, $\mathcal V(x,y)=\mathcal V(x-y)$, even, of range $d$, with the wires coupled over a window $(-L,L)$, $d\ll L$, one finds exactly
\begin{gather}
V_{++}(k)=\frac{e^{2ikL}\,\hat{\mathcal V}_+(-k)-\hat{\mathcal V}_+(k)}{ik},
\nonumber\\
V_{+-}(k)=-i\,\frac{d\hat{\mathcal V}_+}{dk},
\qquad
\hat{\mathcal V}_+(k)=\int_0^\infty e^{ikv}\,\mathcal V(v)\,dv,
\end{gather}
with the remaining components fixed by mirror symmetry. Averaging over the fast phase $2kL$,
\begin{equation}
\big\langle\mathcal J(k)\big\rangle
=\frac{4|\hat{\mathcal V}_+(k)|^2}{k^2}
-2\bigg|\frac{d\hat{\mathcal V}_+}{dk}\bigg|^2
\longrightarrow
\begin{cases}
\dfrac{\hat{\mathcal V}(0)^2}{k^2}, & kd\ll1,\\[8pt]
\dfrac{2\mathcal V(0)^2}{k^4}, & kd\gg1.
\end{cases}
\label{eq:Jwire}
\end{equation}
Here $k=\omega/v_F\sim L_T^{-1}$ is the thermal wavevector of the rectified fluctuation, not the Fermi momentum; the $d$ dependence at $kd\ll1$ resides entirely in $\hat{\mathcal V}(0)\propto\ln(\lambda_s/d)$ (with $\lambda_s$ the gate-screening length), while for $kd\gg1$ the kernel carries $\mathcal V^2(0)\propto1/d^2$ explicitly. In the regime $L_T\gg d$ relevant to experiments the drag thus depends on the interaction only through its zero-momentum component $\hat{\mathcal V}(0)$, the result is manifestly finite (no factor of the wire length survives), and inserting the $kd\ll1$ limit into Eq.~\eqref{eq:master} with the small-$\omega$ asymptotics of Eq.~\eqref{eq:Aclosed} yields Eq.~\eqref{eq:gdmain} with the stated constant. All elements of this chain were validated numerically (Fig.~\ref{fig:kernel}). Beyond the $T^2$ window, Eq.~\eqref{eq:Jwire} predicts a softened, approximately linear-in-$T$ growth for $v_F/d\ll T\ll\Delta_1$ and saturation above $\Delta_1$, as borne out in Fig.~\ref{fig:kernel}(b).

The exact kernel also answers how sensitive the drag is to the relative placement of the two constrictions. A transverse displacement only changes $d$ and is covered by the $d$ dependence above. A longitudinal offset $x_0$ of the bottleneck of wire 2 relative to that of wire 1 enters Eq.~\eqref{eq:kernel} because the sides $s,s'$ are defined relative to each bottleneck: in bottleneck-centered coordinates, $\mathcal V(x,y)\to\mathcal V(x-y-x_0)$. For $d\ll|x_0|\ll L$ and $kd\ll1$ the quadrant integrals can be done in closed form. The cross-side component acquires a nonoscillating piece, $V_{+-}=\hat{\mathcal V}(0)\,x_0\,e^{ikx_0}$ for $x_0>0$ (and $V_{-+}$ for $x_0<0$), so that after phase averaging
\begin{gather}
\langle\mathcal J\rangle=\hat{\mathcal V}(0)^2\Big[\frac1{k^2}-x_0^2\Big],
\qquad
\langle\mathcal J_0\rangle=-\sgn(x_0)\,\hat{\mathcal V}(0)^2x_0^2,
\nonumber\\
\eta_c=\frac{\mathcal J_0}{\mathcal J}=-\frac{\sgn(x_0)\,(kx_0)^2}{1-(kx_0)^2}.
\label{eq:Jshift}
\end{gather}
 After the frequency integration with the thermal weight $\omega^2\sinh^{-2}(\omega/2T)$, for which $\langle\omega^2\rangle=(4\pi^2/5)T^2$, the reciprocal drag and the coupling asymmetry of Sec.~\ref{sec:nonreciprocal} become
\begin{gather}
g_D\to g_D\Big[1-\frac{4\pi^2}{5}\Big(\frac{x_0}{L_T}\Big)^2\Big],
\nonumber\\
\eta_c\simeq-\sgn(x_0)\,\frac{4\pi^2}{5}\Big(\frac{x_0}{L_T}\Big)^2 .
\label{eq:etashift}
\end{gather}
Two conclusions follow. The reciprocal drag and all of its oscillations are insensitive to a longitudinal offset of the constrictions as long as $x_0\ll L_T$ ($L_T\approx10\,\mu$m at $100$\,mK for $v_F=1.5\times10^5$\,m/s), which is the same condition, $d\ll L_T$, under which Eq.~\eqref{eq:gdmain} holds; likewise, the position of the bottlenecks within the coupled window is immaterial after phase averaging as long as both ends of the window are farther than $L_T$ away. An offset is, at the same time, the simplest geometric source of the coupling asymmetry $\eta_c$ required for the field-odd linear drag of Eq.~\eqref{eq:flagship}: $\eta_c$ is odd in $x_0$, as mirror symmetry demands, grows as $(x_0/L_T)^2$, and is of order unity for $x_0\sim0.3L_T$, i.e., for micron offsets at $\sim1$\,K. The sign of the field-odd drag therefore reverses when the direction of the offset is reversed. For $x_0\gtrsim L_T$ the same-side coherence is lost and the drag is suppressed; this is the regime in which displaced constrictions cease to act as a coupled pair.

Two caveats delimit the analysis. The exchange pairing of the interaction legs, $\mathcal V(x_1,x_4)\mathcal V(x_2,x_3)$, maps onto an analogous kernel with $e^{ik(sx-s'y)}$, which for short-ranged translationally invariant coupling contains a term extensive in the coupled length; in the circuit formulation the corresponding contribution enters only at second order in the drive voltage~\cite{Levchenko2008prl} and drops from the linear response, but a fully microscopic verification for the distributed geometry remains worthwhile. Second, the static-interaction approximation is relaxed in Sec.~\ref{sec:circuit}, where the frequency structure of the circuit coupling is treated explicitly.

\bibliography{biblioDragQWQPC}

\end{document}